\documentclass[aps,twocolumn,superscriptaddress,showkeys,showpacs]{revtex4-2}

\usepackage{amsmath}
\usepackage{mathtools} 
\usepackage{amssymb}
\usepackage{amsthm}
\usepackage{bm,upgreek} 
\usepackage{upgreek} 
\usepackage{xcolor}
\usepackage{graphicx}
\usepackage{epstopdf}
\usepackage[colorlinks,linkcolor=blue,anchorcolor=blue,citecolor=blue,urlcolor=blue]{hyperref}

\newcommand{\nn}{{\nonumber}}
\newcommand{\bea}{\begin{eqnarray}}
\newcommand{\eea}{\end{eqnarray}}
\newcommand{\ie}{\textit{i.e.}}

\newcommand{\av}[1]{\left\langle #1 \right\rangle}

\newcommand{\md}{\mathrm{d}}
\newcommand{\me}{\mathrm{e}}
\newcommand{\w}{\omega}

\begin{document}
\title{Quantized bound states around a vortex in anisotropic superconductors}
\author{Ke Xiang}
\affiliation{National Laboratory of Solid State Microstructures $\&$ School of Physics, Nanjing University, Nanjing, 210093, China}
\author{Da Wang} \email{ dawang@nju.edu.cn}
\affiliation{National Laboratory of Solid State Microstructures $\&$ School of Physics, Nanjing University, Nanjing, 210093, China}
\affiliation{Collaborative Innovation Center of Advanced Microstructures, Nanjing 210093, China}
\author{Qiang-Hua Wang} \email{qhwang@nju.edu.cn}
\affiliation{National Laboratory of Solid State Microstructures $\&$ School of Physics, Nanjing University, Nanjing, 210093, China}
\affiliation{Collaborative Innovation Center of Advanced Microstructures, Nanjing 210093, China}

\begin{abstract}
The bound states around a vortex in anisotropic superconductors is a longstanding yet important issue.
In this work, we develop a variational theory on the basis of the Andreev approximation to obtain the energy levels and wave functions of the low-energy quantized bound states in superconductors with anisotropic pairing on arbitrary Fermi surface.
In the case of circular Fermi surface, the effective Schr\"odinger equation yielding the bound state energies gets back to the theory proposed by Volovik and Kopnin many years ago.
Our generalization here enables us to prove the equidistant energy spectrum inside a vortex in a broader class of superconductors.
More importantly, we are now able to obtain the wave functions of these bound states by projecting the quasiclassical wave function on the eigenmodes of the effective Schr\"odinger equation, going beyond the quasiclassical Eilenberger results, which, as we find, are sensitive to the scattering rate.
For the case of isotropic Fermi surface, the spatial profile of the low-energy local density of states is {dominated} near the vortex center and elongates along the gap antinode directions, in addition to the ubiquitous Friedel oscillation arising from the quantum inteference neglected in the Eilenberger theory.
Moreover, as a consequence of the pairing anisotropy, the quantized wave functions develop a peculiar distribution of winding number, which reduces stepwise towards the vortex center.
Our work provides a flexible way to study the vortex bound states in the future.
\keywords{vortices, bound states, pairing symmetries (other than s-wave)}
\pacs{03.75.Lm, 03.65.Ge, 74.20.Rp}
\end{abstract}

\maketitle

\section{Introduction}

Around a vortex in a type-II superconductor, there are a series of low-energy fermionic bound states as firstly obtained by Caroli, de Gennes and Matricon (CdGM) for isotropic s-wave pairing \cite{CdGM1964,deGennesBook}. The energy levels are given by $E_n=(n+\frac12)\w_0$, and the wave functions are labeled by the integer angular momentum quantum number $n$.
The results have been generalized to chiral superconductors with the pairing function $\Delta_\theta=\Delta\me^{iN\theta}$ where $\theta$ is the azimuthal angle of the momentum \cite{Volovik1999}. Although this angular dependence breaks rotational symmetry, it can be formally eliminated by a gauge transformation. As a consequence, $n$ is still a good quantum number.
For odd $N$, the CdGM spectrum becomes $E_n=n\w_0$, among which the zero energy bound state is a Majorana zero mode \cite{Read-Green2000}.
It turns out that the braiding of the vortices holding Majorana zero mode realizes a nontrivial representation of the non-Abelian permutation group \cite{Ivanov_PRL_2001},
which provides an ideal platform for fault-tolerant topological quantum computation \cite{Kitaev2003,Nayak2008,DasSarma2015}.

However, for an anisotropic superconductor with a general pairing $\Delta_\theta$, the angular momentum is no longer conserved by symmetry. This causes difficulty in theoretical treatment, and even debates on the property of the bound states. For example, the spatial profile of local density of states (LDOS) at low energies was found, within the Eilenberger theory, to extend along the gap minima \cite{Hayashi_PRL_1996}, the ``leaky'' directions anticipated from the momentum-dependent Cooper-pair size $\xi_\theta\sim v_F/|\Delta_\theta|$. The same elongation behavior was also reported in d-wave superconductors by both Eilenberger \cite{Schopohl-Maki1995,Ichioka_PRB_1996} and Bogoliubov-de Gennes (BdG) \cite{Franz_PRL_1998} calculations, although the states here are actually resonant rather than bound in nature \cite{Volovik1993,Franz_PRL_1998}. Recently, the possibility of topological nematic superconductivity in Cu$_x$Bi$_2$Se$_3$ attracted much attention, for which the antinode direction is crucial to identify whether its paring is topological or not \cite{Fu_PRL_2010,Fu_PRB_2014}.
BdG and quasiclassical calculations for the p-wave nematic pairing found the elongations are along the gap antinode directions \cite{Bao2018,Yang2019,Nagai_JPSJ_2014}, and a similar elongation was also found for an anisotropic s-wave pairing \cite{Fang2021}. These appear to be different to the early Eilenberger result \cite{Hayashi_PRL_1996}. Furthermore, from the experimental side, a clear and correct understanding of the elongation behavior is important for the scanning tunneling microscopic (STM) experiments \cite{Hess_PRL_1990,Song_S_2011,Hanaguri_PRB_2012,Kaneko_JPSJ_2012,Du_SR_2015,Wang_PRB_2018,Tao_PRX_2018,Chen_NC_2018,Chen_SA_2018,Yuan_NP_2019,Zhang_PRL_2021,Chen_PRL_2021} to probe the gap antinode directions.

Regarding all these issues, we need an accurate understanding of the bound states (with both energy levels and wave functions) for a general pairing $\Delta_\theta$ around a vortex. The original CdGM solution \cite{CdGM1964} is based on the solvability of the particular BdG differential equation, which is difficult to generalize to other pairings. Gygi and Schl\"uter proposed to solve the BdG differential equation numerically with Fourier-Bessel series expansion \cite{Gygi_PRB_1991}. However, its efficiency actually depends on the circular symmetry and parabolic dispersion of the normal state, and the generalization to general pairings on general Fermi surfaces is not straightforward.
One remarkable progress was achieved by Volovik and Kopnin \cite{Volovik1999,VolovikBook,Volovik_PR_2001,KopninBook}, who discovered the bound state energy levels are still equidistant by Bohr-Sommerfeld quantization of the quasiclassical energy in the Andreev approximation.
However, the rationale behind such a quantization procedure is not clearly elaborated, and more importantly, to our best knowledge, there has been no succedent work in the literature to give the real-space (not quasiclassical) wave functions of these quantized bound states around a vortex in anisotropic superconductors after their pioneer work.

In this paper, we first briefly introduce the quasiclassical Andreev approximation and the Bohr-Sommerfeld quantization for the vortex problem in anisotropic superconductrors. Then we propose a variational theory to obtain the energy levels, and meanwhile to obtain the real-space wave functions of the quantized bound states. Our theory is benchmarked by lattice BdG calculations, and generalized to cases with anisotropic Fermi surface.
As a particular result, for anisotropic pairing on circular Fermi surface, the resulting wave function amplitudes and low-energy LDOS are dominated by elongations along the gap antinode directions, in addition to the ubiquitous Friedel oscillation, while the long tails along the gap minima directions are too weak to be resolved clearly.

To facilitate later discussions and as a concrete example, we assume the pairing $\Delta_\theta$ in the following form
\begin{eqnarray}\label{eq:Dtheta}
\Delta_\theta=(\Delta_0+\Delta_\ell\cos\ell\theta)\me^{iN\theta} ,
\end{eqnarray}
where $\theta$ is the direction of the Fermi momentum $\0k_F$ and $N$ is the winding number describing the chirality in momentum space. Here, $\Delta_0$- and $\Delta_\ell$-terms mix to yield an anisotropic pairing.
Around an isolated vortex, the pairing function acquires spatial dependence such as
\begin{eqnarray} \label{eq:PairProfile}
\Delta(r,\phi;\theta)=\Delta_\theta\me^{i\phi}\tanh\frac{r}{\xi},
\end{eqnarray}
where $r$ and $\phi$ are polar coordinates of the planar position vector $\0r$ relative to the vortex center, and
$\xi$ is the coherence length. This particular $\0r$-dependence is {not necessary} in our theory, but is a good approximation for the spatial profile without loss of essential physics and was also used in the original CdGM work \cite{CdGM1964}.
We also limit ourselves to the two-dimensional case, as the generalization to three dimensions is straightforward, given the conserved momentum in the out-of-plane direction. Clever readers may have noticed that Eq.~\ref{eq:PairProfile} is already written down in a quasiclassical way, which means the spatial variance of the pairing function, order of the coherence length $\xi$, should be much longer than $k_F^{-1}$, corresponding to the quasiclassical condition $k_F\xi\sim E_F/|\Delta_\theta|\gg 1$, which applies to most superconductors and is the condition for our following theory.

\section{Andreev approximation and Bohr-Sommerfeld quantization}

\begin{figure}\centering
\includegraphics[trim = 15 10 30 0 , clip,width=\linewidth]{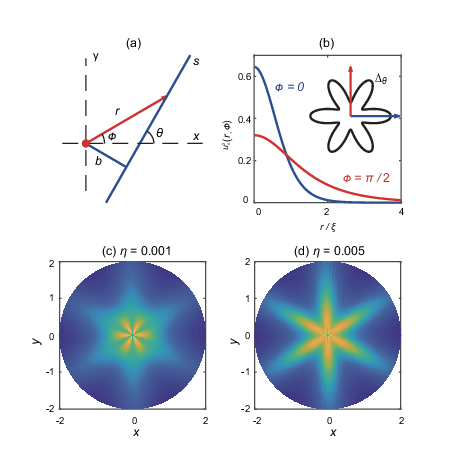}
\caption{ \label{fig:eilenberger}
Quasiclassical Andreev approximation. (a) A quasiclassical path along $\theta$ (direction of $\0k_F$) passing through the spatial point $\0r=(r,\phi)$. The coordinate along the path $s$ and impact number $b$ are directly shown. (b) Wave functions along gap antinode ($\phi=0$) and minima ($\phi=\pi/2$) directions at zero energy with $b=0$, for the six-fold pairing as shown in the inset. (c) and (d) are LDOS (brightness color scale) at zero energy obtained by solving the Eilenberger equation numerically with two different $\eta$.
}
\end{figure}

Under the Andreev approximation, the motions along different $\theta$ are taken to be independent and are described by the quasiclassical wave function \cite{Andreev1964}
\begin{eqnarray}\label{eq:andreev-approx}
\Psi(r,\phi;\theta)= \me^{ik_Fr\cos(\phi-\theta)}\psi(r,\phi;\theta),
\end{eqnarray}
where the fast oscillating factor $\me^{ik_Fr\cos(\phi-\theta)}$ is separated from the slowly varying part $\psi(r,\phi;\theta)$.
The isotropoic Fermi surface is assumed right now, but will be generalized to anisotropic case later.
As shown in Fig.~\ref{fig:eilenberger}(a), we establish the coordinates $(s,b)$ along and perpendicular to the Fermi velocity $\0v_F$ (parallel to $\0k_F$ here), giving $s=r\cos(\phi-\theta)$ and $b=r\sin(\phi-\theta)$.
Substituting $\Psi(r,\phi;\theta)$ into the BdG equation
\begin{eqnarray}
\left[\left(-\frac{\nabla^2}{2m}-\mu\right)\sigma_3+\Delta'\sigma_1-\Delta''\sigma_2\right]\Psi=E\Psi,
\end{eqnarray}
where $\Delta'$ ($\Delta''$) is the real (imaginary) part of $\Delta(r,\phi;\theta)$ and $\sigma_{1,2,3}$ are Pauli matrices,
we obtain the Andreev equation
\begin{eqnarray} \label{eq:andreev}
(-iv_F\partial_s\sigma_3+\Delta'\sigma_1-\Delta''\sigma_2)\psi=E\psi .
\end{eqnarray}

Before proceeding to discuss the quantized bound states, let us first investigate the LDOS $\rho(\w;r,\phi)$, which is an integral over $\theta$ of the quasiclassical LDOS $\rho(\w;r,\phi;\theta)=-\frac{1}{\pi}\mathrm{Tr}[\mathrm{Im}(\sigma_3\hat{g})]$. Here, $\hat{g}(\w;\0r,\theta)$ is the retarded quasiclassical Green's function satisfying the Eilenberger equation \cite{Eilenberger1968}
$-iv_F\partial_s\hat{g}=\left[\sigma_3(E^+\sigma_0+\Delta'\sigma_1-\Delta''\sigma_2),\hat{g}\right]$,
where $E^+=\w+i\eta$ with $\eta$ the scattering rate (or measurement resolution) and $[\cdot,\cdot]$ is the commutation.
Using the Schopohl-Maki parametrization \cite{Schopohl-Maki1995,Schopohl1998}, we reperformed calculations for the six-fold anisotropic pairing as shown in Fig.~\ref{fig:eilenberger}(b), corresponding to $\ell=6$ and $N=0$ in our model Eq.~\ref{eq:Dtheta} with $(\Delta_0,\Delta_6)=(1,0.5)$, as studied in Ref.~\cite{Hayashi_PRL_1996}.
In Figs.~\ref{fig:eilenberger}(c) and (d), we plot the zero energy LDOS with two different values of $\eta$.
For small $\eta=0.001$ in Fig.~\ref{fig:eilenberger}(c), the LDOS shows two different behaviors near and far away from the vortex center. For small $r$, it elongates along the gap antinode directions, consistent with the BdG results for the nematic p-wave \cite{Bao2018,Yang2019} and anisotropic s-wave pairings \cite{Fang2021}.
Instead, for large $\eta=0.005$ in Fig.~\ref{fig:eilenberger}(d), only the elongation  along the gap minima is visible \cite{Hayashi_PRL_1996}.
Therefore, we have found this theoretical dispute in fact comes from different choices of the scattering rate.

One question is raised immediately: how to understand the LDOS difference between near and far field behavior with $\eta\to0$, which is directly given by the quasiclassical wave function $\psi(r,\phi;\theta)$ at zero energy. In the Andreev equation Eq.~\ref{eq:andreev}, after performing a gauge transformation $\psi\to\exp\left(i\frac{1+N}{2}\theta\sigma_3\right)\psi'$ (singular for even $N$), the $\sigma_2$-term vanishes at $s\to\pm\infty$ and thus can be taken as a perturbation. For $b=0$, the $\sigma_2$-term vanishes and the zero energy wave function can be exactly solved as $\psi'=\tilde{\psi}_0[1,-i]^t$ where
\begin{eqnarray}\label{eq:psi0}
\tilde{\psi}_0(r,\theta)=\frac{1}{C_\theta}\exp\left(-\int_0^r\frac{|\Delta(s,0;\theta)|}{v_F}\md s\right) ,
\end{eqnarray}
where $C_\theta$ is a normalization coefficient.
The zero energy LDOS is directly given by $|\tilde{\psi}_0(r,\phi)|^2$ where we have $\phi=\theta$ for the case of $b=0$.
From this solution, $\tilde{\psi}_0$ is most extended along the gap minima directions. Meanwhile, the normalization reduces its amplitude at small $r$ relative to that along the gap antinode directions, as shown in Fig.~\ref{fig:eilenberger}(b). This naturally explains the near-far field difference and the transition occurs at the order of $\xi$.
{Similar results can be obtained for anisotropic Fermi surface. Considering an elliptic Fermi surface with isotropic pairing as an example, from Eq.~\ref{eq:psi0}, smaller $v_F$ (along major axis) provides a more confined (narrow) wave function, while larger $v_F$ (along minor axis) yields a more extended (broad) wave function. Then due to normalization, similar to Fig.~\ref{fig:eilenberger}(b), the total LDOS elongates along major axis for small $r$ and minor axis for large $r$.
}

Next, for $b\ne0$, we can take Eq.~\ref{eq:psi0} as the zeroth order wave function and the $\sigma_2$-term in Eq.~\ref{eq:andreev} (after gauge transformation) as the perturbation, to obtain the first order energy
\begin{align} \label{eq:qcEnergy}
\tilde{E}(b,\theta)=-k_Fb\w_0(\theta),
\end{align}
where $\w_0(\theta)$ is given by
\begin{eqnarray}
\w_0(\theta)=\int_{-\infty}^\infty \frac{|\Delta(r,0;\theta)|}{k_F r}\tilde{\psi}_0^2(r,\theta) \md r .
\end{eqnarray}
Volovik and Kopnin noticed that the quasiclassical energy $\tilde{E}$ can be viewed as an effective Hamiltonian proportional to the angular momentum $L=k_Fb$ conjugate to $\theta$, which is replace by $-i\partial_\theta$ when doing Bohr-Sommerfeld quantization. This procedure naturally gives equidistant quantized bound state energy levels $E_n$ \cite{VolovikBook}. Clearly, such an intuitive procedure is simple and insightful, but it does not tell us how to construct the real-space bound state wave functions, and we are also curious about the rationale behind it.
For these purposes, we propose the following variational theory.

\section{Variational quantization and bound state wave functions}
We make an ansatz that the real-space bound state wave function $\Psi_n(r,\phi)$ can be expressed as a superposition of the quasiclassical wave function $\Psi(r,\phi;\theta)$, which is approximately given by the zeroth order wave function $\psi_0$. Putting the fast oscillation factor and the gauge transformation back together, we have
\begin{eqnarray} &\label{eq:ansatz}
\Psi_n(r,\phi)=\int\md\theta \chi_n(\theta)\me^{ik_Fr\cos(\phi-\theta)}
\nn\\  &\tilde{\psi}_0(r,\theta)\begin{bmatrix}
\me^{i\frac{1+N}{2}\theta} \\
-i\me^{-i\frac{1+N}{2}\theta}
\end{bmatrix},
\end{eqnarray}

where $\chi_n(\theta)$ is to be determined by optimizing the expectation value of the BdG Hamiltonian in such a state. After some lengthy but straightforward algebra (see appendix~\ref{sec:iso} for details), we find $\chi_n(\theta)$ satisfies an effective Schr\"odinger equation
\begin{eqnarray} \label{eq:Heff}
\frac{1}{2}\left\{ -i\frac{\md}{\md\theta}, \w_0(\theta) \right\}\chi_n(\theta)=E_n\chi_n(\theta),
\end{eqnarray}
where $\{\cdot,\cdot\}$ is the anticommutation, which is necessary for Hermiticity.
Actually, Eq.~\ref{eq:Heff} is not others but the quasiclassical energy Eq.~\ref{eq:qcEnergy} with the quantization procedure $L\to-i\partial_\theta$, as a basis in Volovik and Kopnin's pioneer works \cite{Volovik1999,VolovikBook,Volovik_PR_2001,KopninBook}.
Our variational theory here enables us to further establishes Eq.~\ref{eq:ansatz}, which gives the relationship between the quantized wave function $\Psi_n(r,\phi)$ and the quasiclassical one $\Psi(r,\phi;\theta)$. Meanwhile, Eq.~\ref{eq:Heff} is now formally derived rather than an intuitive quantization.

The effective Schr\"odinger equation, Eq.~\ref{eq:Heff}, can be solved as \cite{VolovikBook}
\begin{eqnarray}
\chi_n(\theta)=\frac{C}{\sqrt{\w_0(\theta)}}\exp\left(-iE_n\int_0^\theta\frac{\md\theta'}{\w_0(\theta')}\right),
\end{eqnarray}
where $C$ is a normalization coefficient.
Note that the single-valuedness of $\Psi_n(r,\phi)$ in Eq.~\ref{eq:ansatz} requires $\chi_n$ to be antiperiodic (periodic) in $\theta$ for even (odd) $N$, so $E_n$ is quantized as
$E_n=\left(n+\frac12\right)\tilde{w}_0$
for even $N$, and $E_n=n\tilde{\w}_0$ for odd $N$,
where $n$ is an integer and
$\tilde{\w}_0 =1/\av{\w_0^{-1}(\theta)}$.
The equidistant energy levels were already obtained by Volovik for anisotropic pairings \cite{VolovikBook}.
Our progress is to further obtain the bound state wave functions $\Psi_n(r,\phi)$ via Eq.~\ref{eq:ansatz}.
As a first test, for the isotropic s-wave pairing with $N=0$, we have $\chi_n=\me^{-i(n+\frac12)\theta}$. Substituting $\chi_n$ into Eq.~\ref{eq:ansatz}, using the expansion $\me^{ik_Fr\cos(\phi-\theta)} = \sum_n i^n J_n(k_Fr)\me^{in(\phi-\theta)}$ ($J_n$ is the $n$-th order Bessel function of the first kind) and completing the $\theta$-integral, the CdGM results $\Psi_n=[J_n(k_Fr)\me^{-in\phi},J_{n+1}(k_Fr)\me^{-i(n+1)\phi}]^t\tilde{\psi}_0$ \cite{CdGM1964} can be successfully reproduced.

\begin{figure*}
\centering
\includegraphics[width=\textwidth]{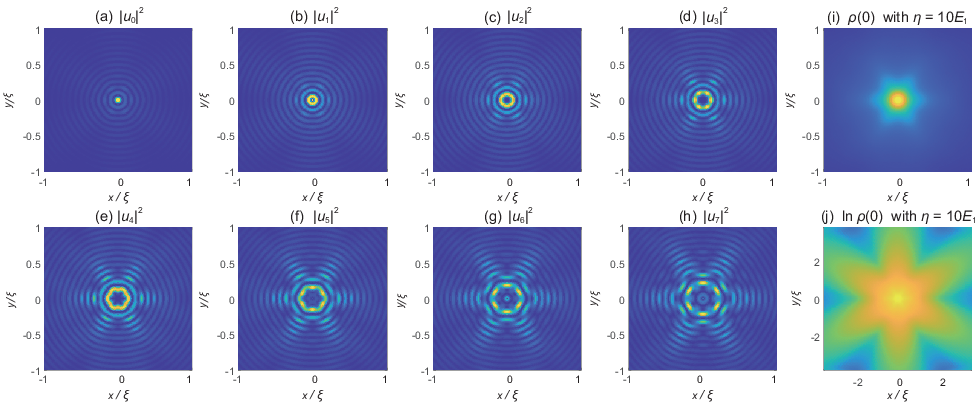}
\caption{ \label{fig:wf6}
The lowest eight quantized bound state wave functions $|u_{0\sim7} |^2$ are plotted (brightness color scale) from (a) to (h), respectively, for the six-fold pairing as shown in Fig.~\ref{fig:eilenberger}(b). The zero energy LDOS are shown in (i) and (j) with linear and logarithmic scales, respectively.
}
\end{figure*}

Now let us apply our theory into the case of six-fold anisotropic pairing studied above. The results with $k_F\xi=40$ are shown in Fig.~\ref{fig:wf6}.
The amplitude square of the lowest eight states $|u_n|^2$, defined in $\Psi_n=[u_n,v_n]^t$, are plotted in Figs.~\ref{fig:wf6}(a-h), respectively.
The six-fold symmetric Friedel oscillation is caused by the quantum interference effect from the coherent superposition in Eq.~\ref{eq:ansatz}, and the wave function amplitude is larger along the gap antinode directions, the same as the near field behavior under the Andreev approximation.
However, the first state $|u_0|^2$ as shown in Fig.~\ref{fig:wf6}(a) shows very weak anisotropic behavior. Therefore, if the experimental resolution is high enough, the zero energy LDOS is expected to show an almost isotropic peak at $r=0$.
Such a feature is quite different from the Andreev approximation which fails at the scale of $r\lesssim k_F^{-1}$.
On the other hand, if the experimental resolution is not sufficiently high, the zero energy LDOS $\rho_0(r,\phi)$ should be contributed by a series of these states, as given by
$\rho_0(r,\phi)=\frac{\eta}{\pi}\sum_n {|u_n(r,\phi)|^2}/{(E_n^2+\eta^2)}$.
We show the results of $\rho_0(r,\phi)$ in Fig.~\ref{fig:wf6}(i) with $\eta=10E_0$, where we see the elongations are along the gap antinode directions.
Instead, the long tails along the gap minima directions at large $r$ is barely visible, unless we plot it in a logarithmic scale as shown in Fig.~\ref{fig:wf6}(j), which in real experiments should be difficult to measure.

\begin{figure}
\includegraphics[width=\linewidth]{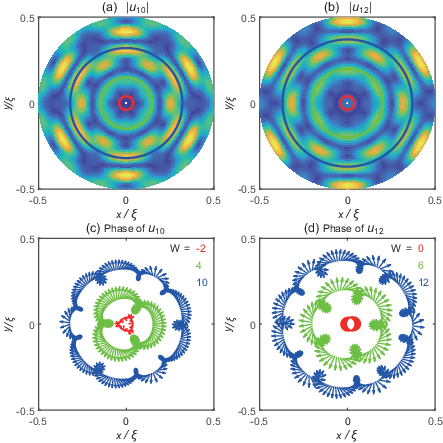}
\caption{ \label{fig:phase6}
The amplitude of the bound state wave functions $u_{10}$ and $u_{12}$ are plotted with brightness color scale in (a) and (b), respectively.
Along three circles indicated by different colors, the phases are represented by arrows and plotted in (c) and (d), respectively, with their winding numbers $W$ as indicated explicitly.
}
\end{figure}

Next, let us examine the phase structure of these bound states.
Since the anisotropic pairing breaks rotational symmetry, the different angular momentum states $\me^{im\phi}$ with $m$ and $m\pm\ell$ are expected to mix together.
For two bound states $u_{10}$ and $u_{12}$, we plot their amplitudes in Figs.~\ref{fig:phase6}(a) and (b). Along three circles (wave function zeros avoided) as indicated by different colors, we plot their phases represented by arrows in Figs.~\ref{fig:phase6}(c) and (d), respectively. As $r$ reduces, the winding number drops from $10$ to $4$ and to $-2$ for $u_{10}$, and from $12$ to $6$ and to $0$ for $u_{12}$, each time by $\ell=6$.
Such a phase structure is in contrast to the isotropic case, where the winding number $W$ does not change with $r$ \cite{CdGM1964}.
Note, however, the winding structure far from the vortex is hard to define as the circular path would be interrupted by zeros caused by the non-circular Friedel oscillation.

\begin{figure*}
\includegraphics[width=\textwidth]{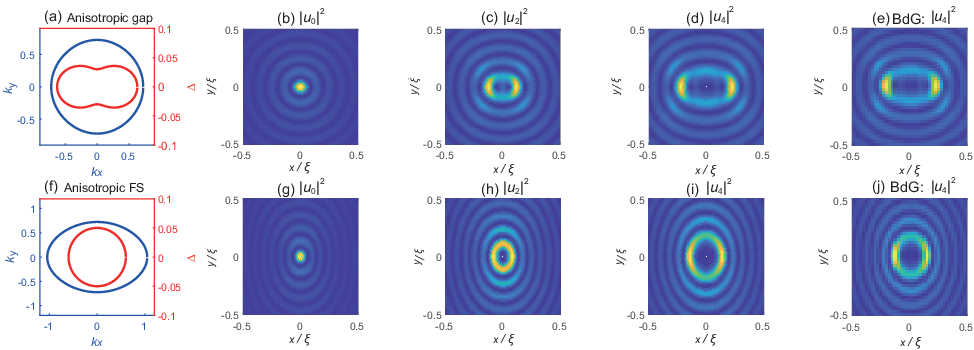}
\caption{ \label{fig:wf2}
For the two-fold anisotropic pairing as shown in (a) with Fermi surface and gap (on the Fermi surface) represented by blue and red colors, the three bound state wave functions $|u_{0,2,4}|^2$ are plotted (brightness color scale) in (b,c,d), respectively. The corresponding result of $|u_4|^2$ obtained by lattice BdG is plotted in (e).
For the two-fold anisotropic Fermi surface with isotropic pairing as shown in (f), the wave functions $|u_{0,2,4}|^2$ are plotted in (g,h,i), respectively. The corresponding BdG result of $|u_{4}|^2$ is given in (j).
}
\end{figure*}

\begin{figure}
\includegraphics[trim=25 3 15 10,clip,width=\linewidth]{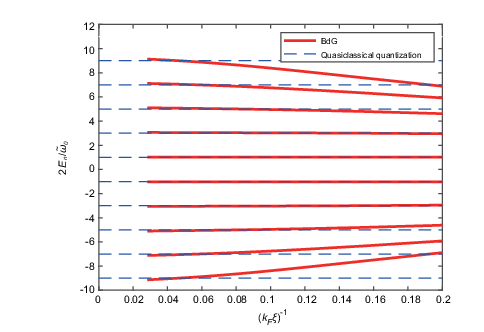}
\caption{ \label{fig:energy}
{In the case of two-fold anisotropic pairing as shown in Fig.~\ref{fig:wf2}(a), the low energy bound state levels are plotted versus $(k_F/\xi)^{-1}$ using both quasiclassical quantization (dashed lines) and BdG (solid lines). }
}
\end{figure}

Before proceeding, we provide a numerical benchmark to our variational theory. For the case of $\ell=2$ as shown in Fig.~\ref{fig:wf2}(a), we plot three states $|u_{0,2,4}|^2$ in Figs.~\ref{fig:wf2}(b,c,d), respectively. Similar to the above case of $\ell=6$, the near field elongations are also along the gap antinode directions. Alternatively, we can model such a pairing on the square lattice with the nearest neighbor hopping $t$, onsite pairing $\Delta_s$ and d-wave pairing $\Delta_d$ on the nearest neighbor bonds. We tune the chemical potential $\mu=-3.5t$ to achieve an essentially circular Fermi surface and we set $(\Delta_s,\Delta_d)=(0.05t,-0.04t)$ to achieve the two-fold symmetric (or nematic) gap as shown in Fig.~\ref{fig:wf2}(a).
On the $400\times400$ lattice with $\xi=40$, we numerically diagonolize the BdG Hamiltonian to obtain the first few eigenstates near the zero energy. Both the energies and wave functions are in good agreement with the quasiclassical quantization results. In Fig.~\ref{fig:wf2}(e), we plot $|u_4|^2$ obtained by BdG, which is in excellent agreement with Fig.~\ref{fig:wf2}(d).
{In Fig.~\ref{fig:energy}, we show the energy levels of the bound states obtained by BdG as varying $(k_F\xi)^{-1}$. Clearly, for large $k_F\xi$ satisfying the quasiclassical condition, the BdG results are in good agreement with the quasiclassical quantization.}

\section{Generalization to anisotropic Fermi surface}

In the above, we assumed a circularly symmetric Fermi surface. The extension to a general anisotropic Fermi surface is as follows. In the ansatz for the wave function, instead of integration over the Fermi angle, we now integrate along the Fermi contour length $\ell$, and $\Psi_n(r,\phi)$ now becomes
\begin{eqnarray} \label{eq:ansatz2}
\Psi_n(r,\phi)=\int\md\ell \chi_n(\ell)\me^{ik_F(\ell)r\cos(\phi-\theta_\ell)}  \nn\\
\tilde{\psi}_0(r,\theta_\ell) \begin{bmatrix}
\me^{i\frac{1+N}{2}\theta_\ell} \\
-i\me^{-i\frac{1+N}{2}\theta_\ell}
\end{bmatrix},
\end{eqnarray}
where both $k_F$ and $\theta$ depend on $\ell$. Then along similar lines as for the isotropic case, we find $\chi_n(\ell)$ satisfies the following effective Schr\"odinger equation (see appendix \ref{sec:aniso} for details)
\begin{eqnarray} \label{eq:Heff2}
\frac{1}{2}\left\{ -i\frac{\md}{\md\ell}, k_F(\ell)\w_0(\ell) \right\}\chi_n(\ell)=E_n\chi_n(\ell),
\end{eqnarray}
subject to a normalization condition
\begin{eqnarray}
\int\frac{\md\ell}{\cos\alpha(\ell)}|\chi_n(\ell)|^2=1,
\end{eqnarray}
where $\alpha(\ell)$ is the angle between the Fermi momentum $\0k_F(\ell)$ and Fermi velocity $\0v_F(\ell)$.
{It is interesting to note that $b$ and $\ell$ are a pair of conjugate variables for the generalized case, as seen from Eq.~\ref{eq:Heff2}.}

Eq.~\ref{eq:ansatz2} and Eq.~\ref{eq:Heff2} are the main results of our present work. They include the above results, Eq.~\ref{eq:ansatz} and Eq.~\ref{eq:Heff}, for the circular Fermi surface as a particular case.
One natural outcome of the anisotropic generalization is that the bound state energies are found to be still quantized equidistantly. The proof is in parallel to the isotropic case following from the single-valuedness of $\chi_n(\ell)$.
Furthermore, the bound state wave functions $\Psi_n$ can be obtained by projecting the quasiclassical wave function $\psi$ on the eigenmodes $\chi_n$ of the effective Schr\"odinger equation as shown in Eq.~\ref{eq:ansatz2}.

As a concrete example, we show the results of a two-fold symmetric Fermi surface with isotropic pairing as shown in Fig.~\ref{fig:wf2}(f). The amplitude square of three wave functions $|u_{0,2,4}|^2$ are plotted in Figs.~\ref{fig:wf2}(g,h,i), respectively. As a benchmark, based on a nematic BdG Hamiltonian on the square lattice with $\mu=-2.5t$, $(t_x,t_y)=(0.5t,t)$ and $\Delta_s=0.05t$, the result of $|u_4|^2$ is shown in Fig.~\ref{fig:wf2}(j), which is in good agreement with Fig.~\ref{fig:wf2}(i). For such a elliptic Fermi surface, the bound state wave pattern elongates along the minor axis, but the splitting of the two peaks (as shown) are along the major axis.

In practice, although the above $\ell$-formalism is universal and can be applied to arbitrary Fermi surface, it is slightly cumbersome to work with $\ell$ as a parameter. To this end, we provide a convenient version if $\theta$ is a \textbf{\textit{1-to-1}} function of $\ell$. In this case, we can still work with $\theta$ instead of $\ell$ as the integration variable. By defining $\chi(\ell)=(\md\theta/\md\ell)\chi(\theta)$, the above $\ell$-formalism can be easily transformed to the $\theta$-formalism, giving rise to
\begin{align}
\Psi_n(r,\phi)=&\int\md\theta \chi_n(\theta)\me^{ik_F(\theta)r\cos(\phi-\theta)} \nn\\
&\tilde{\psi}_0(r,\theta)\begin{bmatrix}
\me^{i\frac{1+N}{2}\theta} \\
-i\me^{-i\frac{1+N}{2}\theta}\end{bmatrix} ,
\end{align}
where $\chi(\theta)$ is determined by
\begin{align}
\frac{1}{2}\left\{ -i\frac{\md}{\md\theta}, \cos\alpha(\theta)\w_0(\theta) \right\}\chi_n(\theta)=E_n\chi_n(\theta) ,
\end{align}
with the normalization condition
\begin{align}
\int\md\theta \frac{1}{k_F(\theta)}\chi_n^*(\theta)\chi_n(\theta)=1 .
\end{align}
In comparison to the isotropic case, $\w_0$ is now replaced by $\w_0 \cos\alpha$, and an additional Jacobian $1/k_F(\theta)$ appears in the normalization condition.

Before closing this paper, we briefly remark on our approach with respect to the Fourier-Bessel series expansion method proposed by Gygi and Schl\"utter \cite{Gygi_PRB_1991}. One advantage of the latter is that the pairing can be solved self-consistently. But it is not easy to apply to general pairing on general Fermi surface, since the BdG differential equation itself is hard to be written down when non-parabolic and non-circular dispersion or pairing exists, and meanwhile its numerical efficiency would be reduced by lack of rotational symmetry.
Instead, our approach is more flexible, which only requires $\0k_F$, $\0v_F$, and $\Delta$ on the Fermi surface, and thus can be broadly applied.

\section{Summary and discussions}

In summary, we have developed a variational theory to obtain the low-energy vortex bound states in superconductors with general pairing on general Fermi surface under the quasiclassical condition $E_F/|\Delta_\theta|\gg1$. In the case of circular Fermi surface, the effective Schr\"odinger equation yielding the bound state energies gets back to the one proposed by Volovik and Kopnin many years ago. But our generalization applies to arbitrary Fermi surface and enables us to prove the equidistantly discrete energy spectrum in a broader class of superconductors. More importantly, we are now able to obtain the wave functions of these bound states by projecting the quasiclassical wave function on the eigenmodes of the effective Schr\"odinger equation.

Our progress makes us to be able to obtain each bound state wave function $\Psi_n$, going beyond the quasiclassical Eilenberger results, which, as we find, are sensitive to the scattering rate. We summarize our findings about the LDOS elongation problem as follows. For isotropic Fermi surface, the spatial profile of the low-energy LDOS elongates along the gap antinode directions near the vortex center at $r\lesssim\xi$, while elongates along the gap minima directions farther but almost invisibly. For the elliptic Fermi surface with isotropic pairing, the elongation directions of the LDOS also change from major to minor axis as going away from the vortex center. Therefore, both pairing and Fermi surface can strongly affect the spatial profile of the LDOS, and thus have to be analyzed carefully in real materials. {In addition to the above, gap modulation, impurity effect, spin-orbit coupling or multi-band character may also have significant effect on the experimental results
\cite{Hess_PRL_1990,Song_S_2011,Hanaguri_PRB_2012,Kaneko_JPSJ_2012,Du_SR_2015,Wang_PRB_2018,Tao_PRX_2018,Chen_NC_2018,Chen_SA_2018,Yuan_NP_2019,Zhang_PRL_2021,Chen_PRL_2021},
}
and are left for future studies.

D.W. thanks N. Schopohl for bringing us Ref.~\cite{Schopohl1998} about technical details in solving the Eilenberger equation.
This work is supported by National Key R\&D Program of China (Grant No. 2022YFA1403201) and National Natural Science Foundation of China (Grant No. 12274205, 12374147, 92365203, and 11874205).

\bibliography{Bibvortex}

\begin{thebibliography}{38}%
\makeatletter
\providecommand \@ifxundefined [1]{%
 \@ifx{#1\undefined}
}%
\providecommand \@ifnum [1]{%
 \ifnum #1\expandafter \@firstoftwo
 \else \expandafter \@secondoftwo
 \fi
}%
\providecommand \@ifx [1]{%
 \ifx #1\expandafter \@firstoftwo
 \else \expandafter \@secondoftwo
 \fi
}%
\providecommand \natexlab [1]{#1}%
\providecommand \enquote  [1]{``#1''}%
\providecommand \bibnamefont  [1]{#1}%
\providecommand \bibfnamefont [1]{#1}%
\providecommand \citenamefont [1]{#1}%
\providecommand \href@noop [0]{\@secondoftwo}%
\providecommand \href [0]{\begingroup \@sanitize@url \@href}%
\providecommand \@href[1]{\@@startlink{#1}\@@href}%
\providecommand \@@href[1]{\endgroup#1\@@endlink}%
\providecommand \@sanitize@url [0]{\catcode `\\12\catcode `\$12\catcode
  `\&12\catcode `\#12\catcode `\^12\catcode `\_12\catcode `\%12\relax}%
\providecommand \@@startlink[1]{}%
\providecommand \@@endlink[0]{}%
\providecommand \url  [0]{\begingroup\@sanitize@url \@url }%
\providecommand \@url [1]{\endgroup\@href {#1}{\urlprefix }}%
\providecommand \urlprefix  [0]{URL }%
\providecommand \Eprint [0]{\href }%
\providecommand \doibase [0]{https://doi.org/}%
\providecommand \selectlanguage [0]{\@gobble}%
\providecommand \bibinfo  [0]{\@secondoftwo}%
\providecommand \bibfield  [0]{\@secondoftwo}%
\providecommand \translation [1]{[#1]}%
\providecommand \BibitemOpen [0]{}%
\providecommand \bibitemStop [0]{}%
\providecommand \bibitemNoStop [0]{.\EOS\space}%
\providecommand \EOS [0]{\spacefactor3000\relax}%
\providecommand \BibitemShut  [1]{\csname bibitem#1\endcsname}%
\let\auto@bib@innerbib\@empty
\bibitem [{\citenamefont {Caroli}\ \emph {et~al.}(1964)\citenamefont {Caroli},
  \citenamefont {De~Gennes},\ and\ \citenamefont {Matricon}}]{CdGM1964}%
  \BibitemOpen
  \bibfield  {author} {\bibinfo {author} {\bibfnamefont {C.}~\bibnamefont
  {Caroli}}, \bibinfo {author} {\bibfnamefont {P.}~\bibnamefont {De~Gennes}},\
  and\ \bibinfo {author} {\bibfnamefont {J.}~\bibnamefont {Matricon}},\
  }\bibfield  {title} {\bibinfo {title} {Bound fermion states on a vortex line
  in a type {II} superconductor},\ }\href
  {https://www.sciencedirect.com/science/article/abs/pii/0031916364903750}
  {\bibfield  {journal} {\bibinfo  {journal} {Phys. Lett.}\ }\textbf {\bibinfo
  {volume} {9}},\ \bibinfo {pages} {307} (\bibinfo {year} {1964})}\BibitemShut
  {NoStop}%
\bibitem [{\citenamefont {de~Gennes}(1999)}]{deGennesBook}%
  \BibitemOpen
  \bibfield  {author} {\bibinfo {author} {\bibfnamefont {P.-G.}\ \bibnamefont
  {de~Gennes}},\ }\href@noop {} {\emph {\bibinfo {title} {Superconductivity of
  Metals and Alloys}}}\ (\bibinfo  {publisher} {{Westview Press}},\ \bibinfo
  {address} {{Boulder}},\ \bibinfo {year} {1999})\BibitemShut {NoStop}%
\bibitem [{\citenamefont {Volovik}(1999)}]{Volovik1999}%
  \BibitemOpen
  \bibfield  {author} {\bibinfo {author} {\bibfnamefont {G.~E.}\ \bibnamefont
  {Volovik}},\ }\bibfield  {title} {\bibinfo {title} {Fermion zero modes on
  vortices in chiral superconductors},\ }\href
  {https://doi.org/10.1134/1.568223} {\bibfield  {journal} {\bibinfo  {journal}
  {JETP Lett.}\ }\textbf {\bibinfo {volume} {70}},\ \bibinfo {pages} {609}
  (\bibinfo {year} {1999})}\BibitemShut {NoStop}%
\bibitem [{\citenamefont {Read}\ and\ \citenamefont
  {Green}(2000)}]{Read-Green2000}%
  \BibitemOpen
  \bibfield  {author} {\bibinfo {author} {\bibfnamefont {N.}~\bibnamefont
  {Read}}\ and\ \bibinfo {author} {\bibfnamefont {D.}~\bibnamefont {Green}},\
  }\bibfield  {title} {\bibinfo {title} {Paired states of fermions in two
  dimensions with breaking of parity and time-reversal symmetries and the
  fractional quantum {Hall} effect},\ }\href
  {https://doi.org/10.1103/PhysRevB.61.10267} {\bibfield  {journal} {\bibinfo
  {journal} {Phys. Rev. B}\ }\textbf {\bibinfo {volume} {61}},\ \bibinfo
  {pages} {10267} (\bibinfo {year} {2000})}\BibitemShut {NoStop}%
\bibitem [{\citenamefont {Ivanov}(2001)}]{Ivanov_PRL_2001}%
  \BibitemOpen
  \bibfield  {author} {\bibinfo {author} {\bibfnamefont {D.~A.}\ \bibnamefont
  {Ivanov}},\ }\bibfield  {title} {\bibinfo {title} {Non-{{Abelian Statistics}}
  of {{Half-Quantum Vortices}} in p-{{Wave Superconductors}}},\ }\href
  {https://doi.org/10.1103/PhysRevLett.86.268} {\bibfield  {journal} {\bibinfo
  {journal} {Phys. Rev. Lett.}\ }\textbf {\bibinfo {volume} {86}},\ \bibinfo
  {pages} {268} (\bibinfo {year} {2001})}\BibitemShut {NoStop}%
\bibitem [{\citenamefont {Kitaev}(2003)}]{Kitaev2003}%
  \BibitemOpen
  \bibfield  {author} {\bibinfo {author} {\bibfnamefont {A.~Y.}\ \bibnamefont
  {Kitaev}},\ }\bibfield  {title} {\bibinfo {title} {Fault-tolerant quantum
  computation by anyons},\ }\href
  {https://doi.org/10.1016/S0003-4916(02)00018-0} {\bibfield  {journal}
  {\bibinfo  {journal} {Ann. Phys.}\ }\textbf {\bibinfo {volume} {303}},\
  \bibinfo {pages} {2} (\bibinfo {year} {2003})}\BibitemShut {NoStop}%
\bibitem [{\citenamefont {Nayak}\ \emph {et~al.}(2008)\citenamefont {Nayak},
  \citenamefont {Stern}, \citenamefont {Freedman},\ and\ \citenamefont
  {Das~Sarma}}]{Nayak2008}%
  \BibitemOpen
  \bibfield  {author} {\bibinfo {author} {\bibfnamefont {C.}~\bibnamefont
  {Nayak}}, \bibinfo {author} {\bibfnamefont {A.}~\bibnamefont {Stern}},
  \bibinfo {author} {\bibfnamefont {M.}~\bibnamefont {Freedman}},\ and\
  \bibinfo {author} {\bibfnamefont {S.}~\bibnamefont {Das~Sarma}},\ }\bibfield
  {title} {\bibinfo {title} {Non-{{Abelian}} anyons and topological quantum
  computation},\ }\href {https://doi.org/10.1103/RevModPhys.80.1083} {\bibfield
   {journal} {\bibinfo  {journal} {Rev. Mod. Phys.}\ }\textbf {\bibinfo
  {volume} {80}},\ \bibinfo {pages} {1083} (\bibinfo {year}
  {2008})}\BibitemShut {NoStop}%
\bibitem [{\citenamefont {Sarma}\ \emph {et~al.}(2015)\citenamefont {Sarma},
  \citenamefont {Freedman},\ and\ \citenamefont {Nayak}}]{DasSarma2015}%
  \BibitemOpen
  \bibfield  {author} {\bibinfo {author} {\bibfnamefont {S.~D.}\ \bibnamefont
  {Sarma}}, \bibinfo {author} {\bibfnamefont {M.}~\bibnamefont {Freedman}},\
  and\ \bibinfo {author} {\bibfnamefont {C.}~\bibnamefont {Nayak}},\ }\bibfield
   {title} {\bibinfo {title} {Majorana zero modes and topological quantum
  computation},\ }\href {https://doi.org/10.1038/npjqi.2015.1} {\bibfield
  {journal} {\bibinfo  {journal} {Npj Quantum Inf.}\ }\textbf {\bibinfo
  {volume} {1}},\ \bibinfo {pages} {1} (\bibinfo {year} {2015})}\BibitemShut
  {NoStop}%
\bibitem [{\citenamefont {Hayashi}\ \emph {et~al.}(1996)\citenamefont
  {Hayashi}, \citenamefont {Ichioka},\ and\ \citenamefont
  {Machida}}]{Hayashi_PRL_1996}%
  \BibitemOpen
  \bibfield  {author} {\bibinfo {author} {\bibfnamefont {N.}~\bibnamefont
  {Hayashi}}, \bibinfo {author} {\bibfnamefont {M.}~\bibnamefont {Ichioka}},\
  and\ \bibinfo {author} {\bibfnamefont {K.}~\bibnamefont {Machida}},\
  }\bibfield  {title} {\bibinfo {title} {Star-{Shaped} {Local} {Density} of
  {States} around {Vortices} in a {Type}-{II} {Superconductor}},\ }\href
  {https://doi.org/10.1103/PhysRevLett.77.4074} {\bibfield  {journal} {\bibinfo
   {journal} {Phys. Rev. Lett.}\ }\textbf {\bibinfo {volume} {77}},\ \bibinfo
  {pages} {4074} (\bibinfo {year} {1996})}\BibitemShut {NoStop}%
\bibitem [{\citenamefont {Schopohl}\ and\ \citenamefont
  {Maki}(1995)}]{Schopohl-Maki1995}%
  \BibitemOpen
  \bibfield  {author} {\bibinfo {author} {\bibfnamefont {N.}~\bibnamefont
  {Schopohl}}\ and\ \bibinfo {author} {\bibfnamefont {K.}~\bibnamefont
  {Maki}},\ }\bibfield  {title} {\bibinfo {title} {Quasiparticle spectrum
  around a vortex line in a d-wave superconductor},\ }\href
  {https://doi.org/10.1103/PhysRevB.52.490} {\bibfield  {journal} {\bibinfo
  {journal} {Phys. Rev. B}\ }\textbf {\bibinfo {volume} {52}},\ \bibinfo
  {pages} {490} (\bibinfo {year} {1995})}\BibitemShut {NoStop}%
\bibitem [{\citenamefont {Ichioka}\ \emph {et~al.}(1996)\citenamefont
  {Ichioka}, \citenamefont {Hayashi}, \citenamefont {Enomoto},\ and\
  \citenamefont {Machida}}]{Ichioka_PRB_1996}%
  \BibitemOpen
  \bibfield  {author} {\bibinfo {author} {\bibfnamefont {M.}~\bibnamefont
  {Ichioka}}, \bibinfo {author} {\bibfnamefont {N.}~\bibnamefont {Hayashi}},
  \bibinfo {author} {\bibfnamefont {N.}~\bibnamefont {Enomoto}},\ and\ \bibinfo
  {author} {\bibfnamefont {K.}~\bibnamefont {Machida}},\ }\bibfield  {title}
  {\bibinfo {title} {Vortex structure in d-wave superconductors},\ }\href
  {https://doi.org/10.1103/PhysRevB.53.15316} {\bibfield  {journal} {\bibinfo
  {journal} {Phys. Rev. B}\ }\textbf {\bibinfo {volume} {53}},\ \bibinfo
  {pages} {15316} (\bibinfo {year} {1996})}\BibitemShut {NoStop}%
\bibitem [{\citenamefont {Franz}\ and\ \citenamefont
  {Tešanović}(1998)}]{Franz_PRL_1998}%
  \BibitemOpen
  \bibfield  {author} {\bibinfo {author} {\bibfnamefont {M.}~\bibnamefont
  {Franz}}\ and\ \bibinfo {author} {\bibfnamefont {Z.}~\bibnamefont
  {Tešanović}},\ }\bibfield  {title} {\bibinfo {title} {Self-{Consistent}
  {Electronic} {Structure} of a $d_{x^2-y^2}$ and a $d_{x^2-y^2}+id_{xy}$
  {Vortex}},\ }\href {https://doi.org/10.1103/PhysRevLett.80.4763} {\bibfield
  {journal} {\bibinfo  {journal} {Phys. Rev. Lett.}\ }\textbf {\bibinfo
  {volume} {80}},\ \bibinfo {pages} {4763} (\bibinfo {year}
  {1998})}\BibitemShut {NoStop}%
\bibitem [{\citenamefont {Volovik}(1993)}]{Volovik1993}%
  \BibitemOpen
  \bibfield  {author} {\bibinfo {author} {\bibfnamefont {G.~E.}\ \bibnamefont
  {Volovik}},\ }\bibfield  {title} {\bibinfo {title} {Superconductivity with
  lines of {{GAP}} nodes: Density of states in the vortex},\ }\href
  {http://jetpletters.ru/ps/1189/article_17954.shtml} {\bibfield  {journal}
  {\bibinfo  {journal} {JETP Lett.}\ }\textbf {\bibinfo {volume} {58}},\
  \bibinfo {pages} {25} (\bibinfo {year} {1993})}\BibitemShut {NoStop}%
\bibitem [{\citenamefont {Fu}\ and\ \citenamefont {Berg}(2010)}]{Fu_PRL_2010}%
  \BibitemOpen
  \bibfield  {author} {\bibinfo {author} {\bibfnamefont {L.}~\bibnamefont
  {Fu}}\ and\ \bibinfo {author} {\bibfnamefont {E.}~\bibnamefont {Berg}},\
  }\bibfield  {title} {\bibinfo {title} {Odd-parity topological
  superconductors: theory and application to {Cu$_x$Bi$_2$Se$_3$}},\ }\href
  {https://doi.org/10.1103/PhysRevLett.105.097001} {\bibfield  {journal}
  {\bibinfo  {journal} {Phys. Rev. Lett.}\ }\textbf {\bibinfo {volume} {105}},\
  \bibinfo {pages} {097001} (\bibinfo {year} {2010})}\BibitemShut {NoStop}%
\bibitem [{\citenamefont {Fu}(2014)}]{Fu_PRB_2014}%
  \BibitemOpen
  \bibfield  {author} {\bibinfo {author} {\bibfnamefont {L.}~\bibnamefont
  {Fu}},\ }\bibfield  {title} {\bibinfo {title} {Odd-parity topological
  superconductor with nematic order: {Application} to {Cu$_x$Bi$_2$Se$_3$}},\
  }\href {https://doi.org/10.1103/PhysRevB.90.100509} {\bibfield  {journal}
  {\bibinfo  {journal} {Phys. Rev. B}\ }\textbf {\bibinfo {volume} {90}},\
  \bibinfo {pages} {100509} (\bibinfo {year} {2014})}\BibitemShut {NoStop}%
\bibitem [{\citenamefont {Bao}\ \emph {et~al.}(2018)\citenamefont {Bao},
  \citenamefont {Tang}, \citenamefont {Lu},\ and\ \citenamefont
  {Wang}}]{Bao2018}%
  \BibitemOpen
  \bibfield  {author} {\bibinfo {author} {\bibfnamefont {W.-C.}\ \bibnamefont
  {Bao}}, \bibinfo {author} {\bibfnamefont {Q.-K.}\ \bibnamefont {Tang}},
  \bibinfo {author} {\bibfnamefont {D.-C.}\ \bibnamefont {Lu}},\ and\ \bibinfo
  {author} {\bibfnamefont {Q.-H.}\ \bibnamefont {Wang}},\ }\bibfield  {title}
  {\bibinfo {title} {Visualizing the $d$ vector in a nematic triplet
  superconductor},\ }\href {https://doi.org/10.1103/PhysRevB.98.054502}
  {\bibfield  {journal} {\bibinfo  {journal} {Phys. Rev. B}\ }\textbf {\bibinfo
  {volume} {98}},\ \bibinfo {pages} {054502} (\bibinfo {year}
  {2018})}\BibitemShut {NoStop}%
\bibitem [{\citenamefont {Yang}\ and\ \citenamefont {Wang}(2019)}]{Yang2019}%
  \BibitemOpen
  \bibfield  {author} {\bibinfo {author} {\bibfnamefont {L.}~\bibnamefont
  {Yang}}\ and\ \bibinfo {author} {\bibfnamefont {Q.-H.}\ \bibnamefont
  {Wang}},\ }\bibfield  {title} {\bibinfo {title} {The direction of the
  d-vector in a nematic triplet superconductor},\ }\href
  {https://doi.org/10.1088/1367-2630/ab40d5} {\bibfield  {journal} {\bibinfo
  {journal} {New J. Phys.}\ }\textbf {\bibinfo {volume} {21}},\ \bibinfo
  {pages} {093036} (\bibinfo {year} {2019})}\BibitemShut {NoStop}%
\bibitem [{\citenamefont {Nagai}(2014)}]{Nagai_JPSJ_2014}%
  \BibitemOpen
  \bibfield  {author} {\bibinfo {author} {\bibfnamefont {Y.}~\bibnamefont
  {Nagai}},\ }\bibfield  {title} {\bibinfo {title} {Field-{Angle}-{Dependent}
  {Low}-{Energy} {Excitations} around a {Vortex} in the {Superconducting}
  {Topological} {Insulator} {Cu$_x$Bi$_2$Se$_3$}},\ }\href
  {https://doi.org/10.7566/JPSJ.83.063705} {\bibfield  {journal} {\bibinfo
  {journal} {J. Phys. Soc. Jpn.}\ }\textbf {\bibinfo {volume} {83}},\ \bibinfo
  {pages} {063705} (\bibinfo {year} {2014})}\BibitemShut {NoStop}%
\bibitem [{\citenamefont {Fang}\ \emph {et~al.}(2021)\citenamefont {Fang},
  \citenamefont {Liu},\ and\ \citenamefont {Cui}}]{Fang2021}%
  \BibitemOpen
  \bibfield  {author} {\bibinfo {author} {\bibfnamefont {D.~L.}\ \bibnamefont
  {Fang}}, \bibinfo {author} {\bibfnamefont {J.~S.}\ \bibnamefont {Liu}},\ and\
  \bibinfo {author} {\bibfnamefont {Y.~K.}\ \bibnamefont {Cui}},\ }\bibfield
  {title} {\bibinfo {title} {Vortex images influenced by superconducting gap
  and {Fermi} surface},\ }\href {https://doi.org/10.1016/j.physc.2021.1353963}
  {\bibfield  {journal} {\bibinfo  {journal} {Phys. C Supercond. Its Appl.}\
  }\textbf {\bibinfo {volume} {591}},\ \bibinfo {pages} {1353963} (\bibinfo
  {year} {2021})}\BibitemShut {NoStop}%
\bibitem [{\citenamefont {Hess}\ \emph {et~al.}(1990)\citenamefont {Hess},
  \citenamefont {Robinson},\ and\ \citenamefont {Waszczak}}]{Hess_PRL_1990}%
  \BibitemOpen
  \bibfield  {author} {\bibinfo {author} {\bibfnamefont {H.~F.}\ \bibnamefont
  {Hess}}, \bibinfo {author} {\bibfnamefont {R.~B.}\ \bibnamefont {Robinson}},\
  and\ \bibinfo {author} {\bibfnamefont {J.~V.}\ \bibnamefont {Waszczak}},\
  }\bibfield  {title} {\bibinfo {title} {Vortex-core structure observed with a
  scanning tunneling microscope},\ }\href
  {https://doi.org/10.1103/PhysRevLett.64.2711} {\bibfield  {journal} {\bibinfo
   {journal} {Phys. Rev. Lett.}\ }\textbf {\bibinfo {volume} {64}},\ \bibinfo
  {pages} {2711} (\bibinfo {year} {1990})}\BibitemShut {NoStop}%
\bibitem [{\citenamefont {Song}\ \emph {et~al.}(2011)\citenamefont {Song},
  \citenamefont {Wang}, \citenamefont {Cheng}, \citenamefont {Jiang},
  \citenamefont {Li}, \citenamefont {Zhang}, \citenamefont {Li}, \citenamefont
  {He}, \citenamefont {Wang}, \citenamefont {Jia} \emph
  {et~al.}}]{Song_S_2011}%
  \BibitemOpen
  \bibfield  {author} {\bibinfo {author} {\bibfnamefont {C.~L.}\ \bibnamefont
  {Song}}, \bibinfo {author} {\bibfnamefont {Y.~L.}\ \bibnamefont {Wang}},
  \bibinfo {author} {\bibfnamefont {P.}~\bibnamefont {Cheng}}, \bibinfo
  {author} {\bibfnamefont {Y.~P.}\ \bibnamefont {Jiang}}, \bibinfo {author}
  {\bibfnamefont {W.}~\bibnamefont {Li}}, \bibinfo {author} {\bibfnamefont
  {T.}~\bibnamefont {Zhang}}, \bibinfo {author} {\bibfnamefont
  {Z.}~\bibnamefont {Li}}, \bibinfo {author} {\bibfnamefont {K.}~\bibnamefont
  {He}}, \bibinfo {author} {\bibfnamefont {L.}~\bibnamefont {Wang}}, \bibinfo
  {author} {\bibfnamefont {J.~F.}\ \bibnamefont {Jia}}, \emph {et~al.},\
  }\bibfield  {title} {\bibinfo {title} {Direct observation of nodes and
  twofold symmetry in fese superconductor},\ }\href
  {https://www.science.org/doi/full/10.1126/science.1202226} {\bibfield
  {journal} {\bibinfo  {journal} {Science}\ }\textbf {\bibinfo {volume}
  {332}},\ \bibinfo {pages} {1410} (\bibinfo {year} {2011})}\BibitemShut
  {NoStop}%
\bibitem [{\citenamefont {Hanaguri}\ \emph {et~al.}(2012)\citenamefont
  {Hanaguri}, \citenamefont {Kitagawa}, \citenamefont {Matsubayashi},
  \citenamefont {Mazaki}, \citenamefont {Uwatoko},\ and\ \citenamefont
  {Takagi}}]{Hanaguri_PRB_2012}%
  \BibitemOpen
  \bibfield  {author} {\bibinfo {author} {\bibfnamefont {T.}~\bibnamefont
  {Hanaguri}}, \bibinfo {author} {\bibfnamefont {K.}~\bibnamefont {Kitagawa}},
  \bibinfo {author} {\bibfnamefont {K.}~\bibnamefont {Matsubayashi}}, \bibinfo
  {author} {\bibfnamefont {Y.}~\bibnamefont {Mazaki}}, \bibinfo {author}
  {\bibfnamefont {Y.}~\bibnamefont {Uwatoko}},\ and\ \bibinfo {author}
  {\bibfnamefont {H.}~\bibnamefont {Takagi}},\ }\bibfield  {title} {\bibinfo
  {title} {Scanning tunneling microscopy/spectroscopy of vortices in
  {LiFeAs}},\ }\href
  {https://journals.aps.org/prb/abstract/10.1103/PhysRevB.85.214505} {\bibfield
   {journal} {\bibinfo  {journal} {Phys. Rev. B}\ }\textbf {\bibinfo {volume}
  {85}},\ \bibinfo {pages} {214505} (\bibinfo {year} {2012})}\BibitemShut
  {NoStop}%
\bibitem [{\citenamefont {Kaneko}\ \emph {et~al.}(2012)\citenamefont {Kaneko},
  \citenamefont {Matsuba}, \citenamefont {Hafiz}, \citenamefont {Yamasaki},
  \citenamefont {Kakizaki}, \citenamefont {Nishida}, \citenamefont {Takeya},
  \citenamefont {Hirata}, \citenamefont {Kawakami}, \citenamefont {Mizushima},\
  and\ \citenamefont {Machida}}]{Kaneko_JPSJ_2012}%
  \BibitemOpen
  \bibfield  {author} {\bibinfo {author} {\bibfnamefont {S.-i.}\ \bibnamefont
  {Kaneko}}, \bibinfo {author} {\bibfnamefont {K.}~\bibnamefont {Matsuba}},
  \bibinfo {author} {\bibfnamefont {M.}~\bibnamefont {Hafiz}}, \bibinfo
  {author} {\bibfnamefont {K.}~\bibnamefont {Yamasaki}}, \bibinfo {author}
  {\bibfnamefont {E.}~\bibnamefont {Kakizaki}}, \bibinfo {author}
  {\bibfnamefont {N.}~\bibnamefont {Nishida}}, \bibinfo {author} {\bibfnamefont
  {H.}~\bibnamefont {Takeya}}, \bibinfo {author} {\bibfnamefont
  {K.}~\bibnamefont {Hirata}}, \bibinfo {author} {\bibfnamefont
  {T.}~\bibnamefont {Kawakami}}, \bibinfo {author} {\bibfnamefont
  {T.}~\bibnamefont {Mizushima}},\ and\ \bibinfo {author} {\bibfnamefont
  {K.}~\bibnamefont {Machida}},\ }\bibfield  {title} {\bibinfo {title} {Quantum
  {Limiting} {Behaviors} of a {Vortex} {Core} in an {Anisotropic} {Gap}
  {Superconductor}},\ }\href {https://doi.org/10.1143/JPSJ.81.063701}
  {\bibfield  {journal} {\bibinfo  {journal} {J. Phys. Soc. Jpn.}\ }\textbf
  {\bibinfo {volume} {81}},\ \bibinfo {pages} {063701} (\bibinfo {year}
  {2012})}\BibitemShut {NoStop}%
\bibitem [{\citenamefont {Du}\ \emph {et~al.}(2015)\citenamefont {Du},
  \citenamefont {Fang}, \citenamefont {Wang}, \citenamefont {Li}, \citenamefont
  {Du}, \citenamefont {Yang}, \citenamefont {Zhu},\ and\ \citenamefont
  {Wen}}]{Du_SR_2015}%
  \BibitemOpen
  \bibfield  {author} {\bibinfo {author} {\bibfnamefont {Z.}~\bibnamefont
  {Du}}, \bibinfo {author} {\bibfnamefont {D.}~\bibnamefont {Fang}}, \bibinfo
  {author} {\bibfnamefont {Z.}~\bibnamefont {Wang}}, \bibinfo {author}
  {\bibfnamefont {Y.}~\bibnamefont {Li}}, \bibinfo {author} {\bibfnamefont
  {G.}~\bibnamefont {Du}}, \bibinfo {author} {\bibfnamefont {H.}~\bibnamefont
  {Yang}}, \bibinfo {author} {\bibfnamefont {X.}~\bibnamefont {Zhu}},\ and\
  \bibinfo {author} {\bibfnamefont {H.-H.}\ \bibnamefont {Wen}},\ }\bibfield
  {title} {\bibinfo {title} {Anisotropic {Superconducting} {Gap} and
  {Elongated} {Vortices} with {Caroli}-{de} {Gennes}-{Matricon} {States} in the
  {New} {Superconductor} {Ta$_4$Pd$_3$Te$_{16}$}},\ }\href
  {https://doi.org/10.1038/srep09408} {\bibfield  {journal} {\bibinfo
  {journal} {Sci. Rep.}\ }\textbf {\bibinfo {volume} {5}},\ \bibinfo {pages}
  {9408} (\bibinfo {year} {2015})}\BibitemShut {NoStop}%
\bibitem [{\citenamefont {Wang}\ \emph {et~al.}(2018)\citenamefont {Wang},
  \citenamefont {Zhang}, \citenamefont {Lv}, \citenamefont {Ding},
  \citenamefont {Wang}, \citenamefont {Li}, \citenamefont {He}, \citenamefont
  {Song}, \citenamefont {Ma},\ and\ \citenamefont {Xue}}]{Wang_PRB_2018}%
  \BibitemOpen
  \bibfield  {author} {\bibinfo {author} {\bibfnamefont {W.-L.}\ \bibnamefont
  {Wang}}, \bibinfo {author} {\bibfnamefont {Y.-M.}\ \bibnamefont {Zhang}},
  \bibinfo {author} {\bibfnamefont {Y.-F.}\ \bibnamefont {Lv}}, \bibinfo
  {author} {\bibfnamefont {H.}~\bibnamefont {Ding}}, \bibinfo {author}
  {\bibfnamefont {L.}~\bibnamefont {Wang}}, \bibinfo {author} {\bibfnamefont
  {W.}~\bibnamefont {Li}}, \bibinfo {author} {\bibfnamefont {K.}~\bibnamefont
  {He}}, \bibinfo {author} {\bibfnamefont {C.-L.}\ \bibnamefont {Song}},
  \bibinfo {author} {\bibfnamefont {X.-C.}\ \bibnamefont {Ma}},\ and\ \bibinfo
  {author} {\bibfnamefont {Q.-K.}\ \bibnamefont {Xue}},\ }\bibfield  {title}
  {\bibinfo {title} {Anisotropic superconductivity and elongated vortices with
  unusual bound states in quasi-one-dimensional nickel-bismuth compounds},\
  }\href {https://doi.org/10.1103/PhysRevB.97.134524} {\bibfield  {journal}
  {\bibinfo  {journal} {Phys. Rev. B}\ }\textbf {\bibinfo {volume} {97}},\
  \bibinfo {pages} {134524} (\bibinfo {year} {2018})}\BibitemShut {NoStop}%
\bibitem [{\citenamefont {Tao}\ \emph {et~al.}(2018)\citenamefont {Tao},
  \citenamefont {Yan}, \citenamefont {Liu}, \citenamefont {Wang}, \citenamefont
  {Ando}, \citenamefont {Wang}, \citenamefont {Zhang},\ and\ \citenamefont
  {Feng}}]{Tao_PRX_2018}%
  \BibitemOpen
  \bibfield  {author} {\bibinfo {author} {\bibfnamefont {R.}~\bibnamefont
  {Tao}}, \bibinfo {author} {\bibfnamefont {Y.-J.}\ \bibnamefont {Yan}},
  \bibinfo {author} {\bibfnamefont {X.}~\bibnamefont {Liu}}, \bibinfo {author}
  {\bibfnamefont {Z.-W.}\ \bibnamefont {Wang}}, \bibinfo {author}
  {\bibfnamefont {Y.}~\bibnamefont {Ando}}, \bibinfo {author} {\bibfnamefont
  {Q.-H.}\ \bibnamefont {Wang}}, \bibinfo {author} {\bibfnamefont
  {T.}~\bibnamefont {Zhang}},\ and\ \bibinfo {author} {\bibfnamefont {D.-L.}\
  \bibnamefont {Feng}},\ }\bibfield  {title} {\bibinfo {title} {Direct
  {Visualization} of the {Nematic} {Superconductivity} in
  {Cu$_x$Bi$_2$Se$_3$}},\ }\href {https://doi.org/10.1103/PhysRevX.8.041024}
  {\bibfield  {journal} {\bibinfo  {journal} {Phys. Rev. X}\ }\textbf {\bibinfo
  {volume} {8}},\ \bibinfo {pages} {041024} (\bibinfo {year}
  {2018})}\BibitemShut {NoStop}%
\bibitem [{\citenamefont {Chen}\ \emph
  {et~al.}(2018{\natexlab{a}})\citenamefont {Chen}, \citenamefont {Chen},
  \citenamefont {Yang}, \citenamefont {Du}, \citenamefont {Zhu}, \citenamefont
  {Wang},\ and\ \citenamefont {Wen}}]{Chen_NC_2018}%
  \BibitemOpen
  \bibfield  {author} {\bibinfo {author} {\bibfnamefont {M.}~\bibnamefont
  {Chen}}, \bibinfo {author} {\bibfnamefont {X.}~\bibnamefont {Chen}}, \bibinfo
  {author} {\bibfnamefont {H.}~\bibnamefont {Yang}}, \bibinfo {author}
  {\bibfnamefont {Z.}~\bibnamefont {Du}}, \bibinfo {author} {\bibfnamefont
  {X.}~\bibnamefont {Zhu}}, \bibinfo {author} {\bibfnamefont {E.}~\bibnamefont
  {Wang}},\ and\ \bibinfo {author} {\bibfnamefont {H.-H.}\ \bibnamefont
  {Wen}},\ }\bibfield  {title} {\bibinfo {title} {Discrete energy levels of
  caroli-de gennes-matricon states in quantum limit in
  {FeTe$_{0.55}$Se$_{0.45}$}},\ }\href
  {https://doi.org/10.1038/s41467-018-03404-8} {\bibfield  {journal} {\bibinfo
  {journal} {Nat. Commun.}\ }\textbf {\bibinfo {volume} {9}},\ \bibinfo {pages}
  {970} (\bibinfo {year} {2018}{\natexlab{a}})}\BibitemShut {NoStop}%
\bibitem [{\citenamefont {Chen}\ \emph
  {et~al.}(2018{\natexlab{b}})\citenamefont {Chen}, \citenamefont {Chen},
  \citenamefont {Yang}, \citenamefont {Du},\ and\ \citenamefont
  {Wen}}]{Chen_SA_2018}%
  \BibitemOpen
  \bibfield  {author} {\bibinfo {author} {\bibfnamefont {M.}~\bibnamefont
  {Chen}}, \bibinfo {author} {\bibfnamefont {X.}~\bibnamefont {Chen}}, \bibinfo
  {author} {\bibfnamefont {H.}~\bibnamefont {Yang}}, \bibinfo {author}
  {\bibfnamefont {Z.}~\bibnamefont {Du}},\ and\ \bibinfo {author}
  {\bibfnamefont {H.~H.}\ \bibnamefont {Wen}},\ }\bibfield  {title} {\bibinfo
  {title} {Superconductivity with twofold symmetry in
  {Bi$_2$Te$_3$/FeTe$_{0.55}$Se$_{0.45}$} heterostructures},\ }\href
  {https://doi.org/10.1126/sciadv.aat1084} {\bibfield  {journal} {\bibinfo
  {journal} {Sci. Adv.}\ }\textbf {\bibinfo {volume} {4}},\ \bibinfo {pages}
  {eaat1084} (\bibinfo {year} {2018}{\natexlab{b}})}\BibitemShut {NoStop}%
\bibitem [{\citenamefont {Yuan}\ \emph {et~al.}(2019)\citenamefont {Yuan},
  \citenamefont {Pan}, \citenamefont {Wang}, \citenamefont {Fang},
  \citenamefont {Song}, \citenamefont {Wang}, \citenamefont {He}, \citenamefont
  {Ma}, \citenamefont {Zhang}, \citenamefont {Huang}, \citenamefont {Li},\ and\
  \citenamefont {Xue}}]{Yuan_NP_2019}%
  \BibitemOpen
  \bibfield  {author} {\bibinfo {author} {\bibfnamefont {Y.}~\bibnamefont
  {Yuan}}, \bibinfo {author} {\bibfnamefont {J.}~\bibnamefont {Pan}}, \bibinfo
  {author} {\bibfnamefont {X.}~\bibnamefont {Wang}}, \bibinfo {author}
  {\bibfnamefont {Y.}~\bibnamefont {Fang}}, \bibinfo {author} {\bibfnamefont
  {C.}~\bibnamefont {Song}}, \bibinfo {author} {\bibfnamefont {L.}~\bibnamefont
  {Wang}}, \bibinfo {author} {\bibfnamefont {K.}~\bibnamefont {He}}, \bibinfo
  {author} {\bibfnamefont {X.}~\bibnamefont {Ma}}, \bibinfo {author}
  {\bibfnamefont {H.}~\bibnamefont {Zhang}}, \bibinfo {author} {\bibfnamefont
  {F.}~\bibnamefont {Huang}}, \bibinfo {author} {\bibfnamefont
  {W.}~\bibnamefont {Li}},\ and\ \bibinfo {author} {\bibfnamefont {Q.-K.}\
  \bibnamefont {Xue}},\ }\bibfield  {title} {\bibinfo {title} {Evidence of
  anisotropic majorana bound states in {2M-WS$_2$}},\ }\href
  {https://doi.org/10.1038/s41567-019-0576-7} {\bibfield  {journal} {\bibinfo
  {journal} {Nat. Phys.}\ }\textbf {\bibinfo {volume} {15}},\ \bibinfo {pages}
  {1046} (\bibinfo {year} {2019})}\BibitemShut {NoStop}%
\bibitem [{\citenamefont {Zhang}\ \emph {et~al.}(2021)\citenamefont {Zhang},
  \citenamefont {Bao}, \citenamefont {Chen}, \citenamefont {Li}, \citenamefont
  {Lu}, \citenamefont {Hu}, \citenamefont {Yang}, \citenamefont {Zhao},
  \citenamefont {Yan}, \citenamefont {Dong}, \citenamefont {Wang},
  \citenamefont {Zhang},\ and\ \citenamefont {Feng}}]{Zhang_PRL_2021}%
  \BibitemOpen
  \bibfield  {author} {\bibinfo {author} {\bibfnamefont {T.}~\bibnamefont
  {Zhang}}, \bibinfo {author} {\bibfnamefont {W.}~\bibnamefont {Bao}}, \bibinfo
  {author} {\bibfnamefont {C.}~\bibnamefont {Chen}}, \bibinfo {author}
  {\bibfnamefont {D.}~\bibnamefont {Li}}, \bibinfo {author} {\bibfnamefont
  {Z.}~\bibnamefont {Lu}}, \bibinfo {author} {\bibfnamefont {Y.}~\bibnamefont
  {Hu}}, \bibinfo {author} {\bibfnamefont {W.}~\bibnamefont {Yang}}, \bibinfo
  {author} {\bibfnamefont {D.}~\bibnamefont {Zhao}}, \bibinfo {author}
  {\bibfnamefont {Y.}~\bibnamefont {Yan}}, \bibinfo {author} {\bibfnamefont
  {X.}~\bibnamefont {Dong}}, \bibinfo {author} {\bibfnamefont {Q.-H.}\
  \bibnamefont {Wang}}, \bibinfo {author} {\bibfnamefont {T.}~\bibnamefont
  {Zhang}},\ and\ \bibinfo {author} {\bibfnamefont {D.}~\bibnamefont {Feng}},\
  }\bibfield  {title} {\bibinfo {title} {Observation of {Distinct} {Spatial}
  {Distributions} of the {Zero} and {Nonzero} {Energy} {Vortex} {Modes} in
  {Li$_{0.84}$Fe$_{0.16}$OHFeSe}},\ }\href
  {https://doi.org/10.1103/PhysRevLett.126.127001} {\bibfield  {journal}
  {\bibinfo  {journal} {Phys. Rev. Lett.}\ }\textbf {\bibinfo {volume} {126}},\
  \bibinfo {pages} {127001} (\bibinfo {year} {2021})}\BibitemShut {NoStop}%
\bibitem [{\citenamefont {Chen}\ \emph {et~al.}(2021)\citenamefont {Chen},
  \citenamefont {Duan}, \citenamefont {Fan}, \citenamefont {Hong},
  \citenamefont {Chen}, \citenamefont {Yang}, \citenamefont {Li}, \citenamefont
  {Luo},\ and\ \citenamefont {Wen}}]{Chen_PRL_2021}%
  \BibitemOpen
  \bibfield  {author} {\bibinfo {author} {\bibfnamefont {X.}~\bibnamefont
  {Chen}}, \bibinfo {author} {\bibfnamefont {W.}~\bibnamefont {Duan}}, \bibinfo
  {author} {\bibfnamefont {X.}~\bibnamefont {Fan}}, \bibinfo {author}
  {\bibfnamefont {W.}~\bibnamefont {Hong}}, \bibinfo {author} {\bibfnamefont
  {K.}~\bibnamefont {Chen}}, \bibinfo {author} {\bibfnamefont {H.}~\bibnamefont
  {Yang}}, \bibinfo {author} {\bibfnamefont {S.}~\bibnamefont {Li}}, \bibinfo
  {author} {\bibfnamefont {H.}~\bibnamefont {Luo}},\ and\ \bibinfo {author}
  {\bibfnamefont {H.-H.}\ \bibnamefont {Wen}},\ }\bibfield  {title} {\bibinfo
  {title} {Friedel {Oscillations} of {Vortex} {Bound} {States} under {Extreme}
  {Quantum} {Limit} in {KCa$_2$Fe$_4$As$_4$F$_2$}},\ }\href
  {https://doi.org/10.1103/PhysRevLett.126.257002} {\bibfield  {journal}
  {\bibinfo  {journal} {Phys. Rev. Lett.}\ }\textbf {\bibinfo {volume} {126}},\
  \bibinfo {pages} {257002} (\bibinfo {year} {2021})}\BibitemShut {NoStop}%
\bibitem [{\citenamefont {Gygi}\ and\ \citenamefont
  {Schlüter}(1991)}]{Gygi_PRB_1991}%
  \BibitemOpen
  \bibfield  {author} {\bibinfo {author} {\bibfnamefont {F.}~\bibnamefont
  {Gygi}}\ and\ \bibinfo {author} {\bibfnamefont {M.}~\bibnamefont
  {Schlüter}},\ }\bibfield  {title} {\bibinfo {title} {Self-consistent
  electronic structure of a vortex line in a type-{II} superconductor},\ }\href
  {https://doi.org/10.1103/PhysRevB.43.7609} {\bibfield  {journal} {\bibinfo
  {journal} {Phys. Rev. B}\ }\textbf {\bibinfo {volume} {43}},\ \bibinfo
  {pages} {7609} (\bibinfo {year} {1991})}\BibitemShut {NoStop}%
\bibitem [{\citenamefont {Volovik}(2003)}]{VolovikBook}%
  \BibitemOpen
  \bibfield  {author} {\bibinfo {author} {\bibfnamefont {G.~E.}\ \bibnamefont
  {Volovik}},\ }\href@noop {} {\emph {\bibinfo {title} {The Universe in a
  Helium Droplet}}}\ (\bibinfo  {publisher} {{Oxford University Press}},\
  \bibinfo {address} {{Oxford}},\ \bibinfo {year} {2003})\BibitemShut {NoStop}%
\bibitem [{\citenamefont {Volovik}(2001)}]{Volovik_PR_2001}%
  \BibitemOpen
  \bibfield  {author} {\bibinfo {author} {\bibfnamefont {G.~E.}\ \bibnamefont
  {Volovik}},\ }\bibfield  {title} {\bibinfo {title} {Superfluid analogies of
  cosmological phenomena},\ }\href
  {https://doi.org/10.1016/S0370-1573(00)00139-3} {\bibfield  {journal}
  {\bibinfo  {journal} {Phys. Rep.}\ }\textbf {\bibinfo {volume} {351}},\
  \bibinfo {pages} {195} (\bibinfo {year} {2001})}\BibitemShut {NoStop}%
\bibitem [{\citenamefont {Kopnin}(2001)}]{KopninBook}%
  \BibitemOpen
  \bibfield  {author} {\bibinfo {author} {\bibfnamefont {N.~B.}\ \bibnamefont
  {Kopnin}},\ }\href@noop {} {\emph {\bibinfo {title} {Theory of nonequilibrium
  superconductivity}}}\ (\bibinfo  {publisher} {Oxford University Press},\
  \bibinfo {address} {{Oxford}},\ \bibinfo {year} {2001})\BibitemShut {NoStop}%
\bibitem [{\citenamefont {Andreev}(1964)}]{Andreev1964}%
  \BibitemOpen
  \bibfield  {author} {\bibinfo {author} {\bibfnamefont {A.~F.}\ \bibnamefont
  {Andreev}},\ }\bibfield  {title} {\bibinfo {title} {The thermal conductivity
  of the intermediate state in superconductors},\ }\href
  {http://jetp.ras.ru/cgi-bin/e/index/e/19/5/p1228?a=list} {\bibfield
  {journal} {\bibinfo  {journal} {Sov. Phys. JETP}\ }\textbf {\bibinfo {volume}
  {46}},\ \bibinfo {pages} {1823} (\bibinfo {year} {1964})}\BibitemShut
  {NoStop}%
\bibitem [{\citenamefont {Eilenberger}(1968)}]{Eilenberger1968}%
  \BibitemOpen
  \bibfield  {author} {\bibinfo {author} {\bibfnamefont {G.}~\bibnamefont
  {Eilenberger}},\ }\bibfield  {title} {\bibinfo {title} {Transformation of
  {{Gorkov}}'s equation for type {{II}} superconductors into transport-like
  equations},\ }\href {https://doi.org/10.1007/BF01379803} {\bibfield
  {journal} {\bibinfo  {journal} {Z. Physik}\ }\textbf {\bibinfo {volume}
  {214}},\ \bibinfo {pages} {195} (\bibinfo {year} {1968})}\BibitemShut
  {NoStop}%
\bibitem [{\citenamefont {Schopohl}(1998)}]{Schopohl1998}%
  \BibitemOpen
  \bibfield  {author} {\bibinfo {author} {\bibfnamefont {N.}~\bibnamefont
  {Schopohl}},\ }\bibfield  {title} {\bibinfo {title} {Transformation of the
  {Eilenberger} {Equations} of {Superconductivity} to a {Scalar} {Riccati}
  {Equation}},\ }\href {http://arxiv.org/abs/cond-mat/9804064} {\bibfield
  {journal} {\bibinfo  {journal} {arXiv:9804064}\ } (\bibinfo {year}
  {1998})}\BibitemShut {NoStop}%
\end{thebibliography}%

\appendix
\begin{widetext}
\section{Isotropic Fermi surface}\label{sec:iso}
In this appendix, we derive the effective Schr\"odinger equation for the case of isotropic Fermi surface, and then generalize it to the case of anisotropic Fermi surface in the next section.
Under the quasiclassical approximation, we  write the bound state wave function $\Psi(r,\phi)$ as
\begin{align} \label{eq:ansatz-sm}
\Psi(r,\phi)=\int\md\theta \chi(\theta) \me^{ik_Fr\cos(\phi-\theta)}\psi(r,\phi;\theta) ,
\end{align}
where $\chi(\theta)$ is to be determined. If we are interested in the low-energy bound states, we can take $\psi(r,\phi;\theta)$ to be the zeroth order wave function $\psi_0(r,\theta)$ explicitly given by \cite{VolovikBook}
\begin{align} \label{eq:approx}
\psi(r,\phi;\theta)\approx\psi_0(r,\theta)=\underbrace{ \frac{1}{C_\theta}\exp\left(-\int_0^r\frac{|\Delta(s,0;\theta)|}{v_F}\md s\right) }_{\tilde{\psi}_0} \begin{bmatrix}
\me^{i\frac{1+N}{2}\theta} \\
-i\me^{-i\frac{1+N}{2}\theta}
\end{bmatrix} .
\end{align}
Eq.~\ref{eq:ansatz-sm} is our ansatz Eq.~\ref{eq:ansatz} in the main text.
The approximation in Eq.~\ref{eq:approx} is explained in  Fig.~\ref{fig:psi0}. For any spatial point $(r,\phi)$ (blue or green points), the quasiclassical wave function $\psi(r,\phi;\theta)$ is approximated by the wave functions at point 1 or 2 both with the zeroth order wave function $\psi_0(r,\theta)$. The error in this approximation is of order $b=r\sin(\phi-\theta)$, and would cause higher order corrections to the low-energy (near zero) $\tilde{E}=-k_Fb\w_0(\theta)$ which is already first order in $b$.
In this sense, the approximation is well justified.

\begin{figure}[h]
\includegraphics[width=0.35\linewidth]{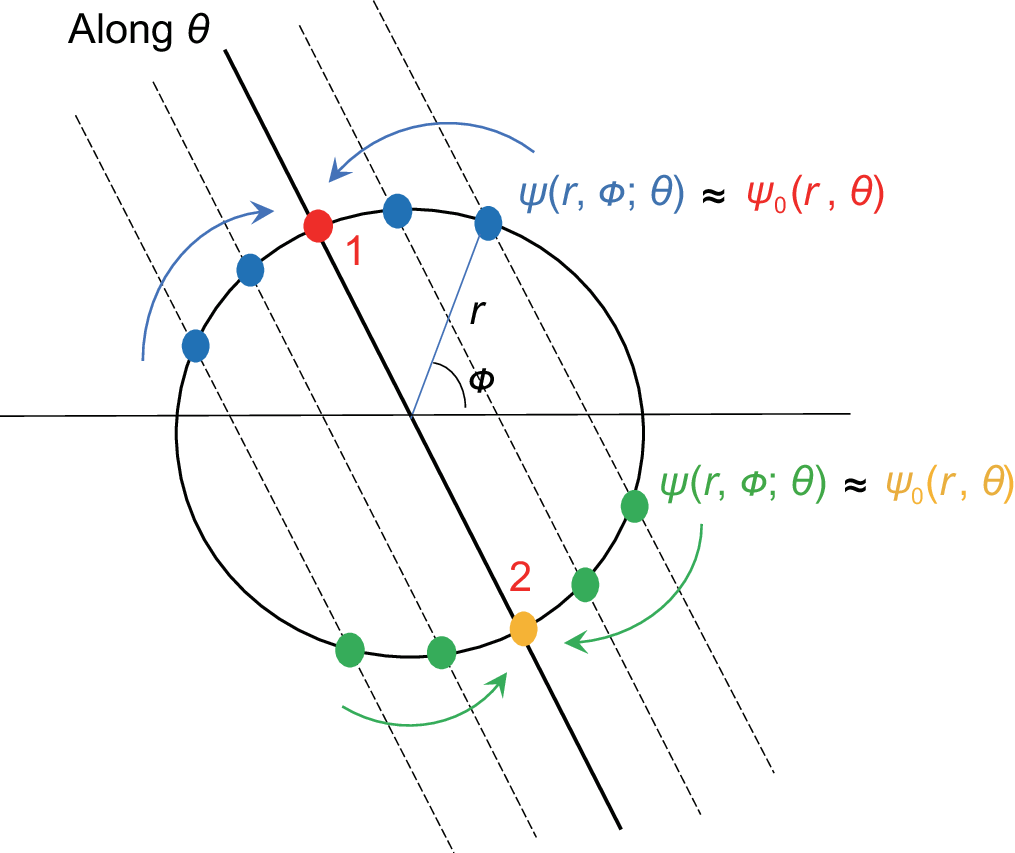}
\caption{\label{fig:psi0}
A scheme of a series of quasiclassical paths along $\theta$. The wave functions $\psi(r,\phi;\theta)$ on the circle with radius $r$ (blue and green dots) are approximated by the wave function $\psi_0(r,\theta)$ at the two points (1 and 2) on the path through the center.
}
\end{figure}

In the following, we vary the energy $E=\av{\Psi|H|\Psi}$ with respect to $\chi(\theta)$, where $H$ is the BdG Hamiltonian. Explicitly, we have
\begin{align}
E&=\frac12\int\md\theta' \int\md\theta \int \md^2\0r \chi^*(\theta') \psi_0^\dag(r,\theta') \me^{-ik_Fr\cos(\phi-\theta')} \me^{ik_Fr\cos(\phi-\theta)} \overrightarrow{H}_A(\theta)  \psi_0(r,\theta)\chi(\theta) \nn\\
&\qquad\qquad\qquad\qquad +\chi^*(\theta') \psi_0^\dag(r,\theta')\overleftarrow{H}_A(\theta') \me^{-ik_Fr\cos(\phi-\theta')} \me^{ik_Fr\cos(\phi-\theta)}  \psi_0(r,\theta)\chi(\theta) ,
\end{align}
where $H_A(\theta)$ is the quasiclassical Andreev Hamiltonian defined on the quasiclassical path along $\theta$ with the impact number $b=r\sin(\phi-\theta)$. Here, symmetrization has been performed since $H_A$ can act on either left or right side, indicated by the overhead arrows.
Since $\psi_0(r,\theta)$ is the zeroth order eigen function of $H_A(\theta)$, it can be replaced by the first order eigen energy $\tilde{E}(\theta,b)=-k_Fb\w_0(\theta)$ where $\w_0$ is given by \cite{VolovikBook}
\begin{align}
\w_0(\theta)=\int_{-\infty}^\infty \frac{|\Delta(r,0;\theta)|}{k_F r}\tilde{\psi}_0^2(r,\theta) \md r .
\end{align}

Our key observation is the following replacement of $k_Fb$ by $-i\partial_\theta$ when acting on the fast oscillation factor, \ie
\begin{align}
k_Fb\me^{ik_Fr\cos(\phi-\theta)}=k_Fr\sin(\phi-\theta)\me^{ik_Fr\cos(\phi-\theta)} = -i\partial_\theta \me^{ik_Fr\cos(\phi-\theta)} .
\end{align}
Then, the total energy $E$ becomes
\begin{align}
E&=\frac12\int\md\theta' \int\md\theta \int \md^2\0r \chi^*(\theta') \psi_0^\dag(r,\theta') \me^{-ik_Fr\cos(\phi-\theta')}  \psi_0(r,\theta)\chi(\theta) \w_0(\theta) \left[i\partial_\theta \me^{ik_Fr\cos(\phi-\theta)} \right] \nn\\
&\qquad\qquad\qquad\qquad +\left[-i\partial_{\theta'}  \me^{ik_Fr\cos(\phi-\theta')} \right] \w_0(\theta') \chi^*(\theta') \psi_0^\dag(r,\theta')\me^{ik_Fr\cos(\phi-\theta)} \psi_0(r,\theta) \chi(\theta) \nn\\
&=\frac12\int\md\theta' \int\md\theta \int \md^2\0r \chi^*(\theta') \psi_0^\dag(r,\theta') \me^{-ik_Fr\cos(\phi-\theta')} \me^{ik_Fr\cos(\phi-\theta)} \left[-i\partial_\theta \w_0(\theta) \psi_0(r,\theta) \chi(\theta)\right] \nn\\
&\qquad\qquad\qquad\qquad +\left[i\partial_{\theta'} \w_0(\theta') \chi^*(\theta') \psi_0^\dag(r,\theta') \right] \me^{-ik_Fr\cos(\phi-\theta')} \me^{ik_Fr\cos(\phi-\theta)}  \chi(\theta) \psi_0(r,\theta) .
\end{align}
In the second equality, integration by parts is performed. Next, we expand $\me^{ik_Fr\cos(\phi-\theta)}$ as
\begin{align} \label{eq:bessel-expansion}
\me^{ik_Fr\cos(\phi-\theta)} = \sum_n i^n J_n(k_Fr)\me^{in(\phi-\theta)} ,
\end{align}
to get
\begin{align}
E&=\frac12\int\md\theta' \int\md\theta \int r\md r \int\md\phi \chi^*(\theta') \nn\\
&\quad \psi_0^\dag(r,\theta')
 \sum_n (-i)^n J_n(k_Fr)\me^{-in(\phi-\theta')}
\sum_m i^m J_m(k_Fr)\me^{im(\phi-\theta)}
\left[-i\partial_\theta \w_0(\theta) \psi_0(r,\theta) \chi(\theta)\right] \nn\\
&+\left[i\partial_{\theta'} \w_0(\theta') \chi^*(\theta') \psi_0^\dag(r,\theta') \right] \sum_n (-i)^n J_n(k_Fr)\me^{-in(\phi-\theta')}
\sum_m i^m J_m(k_Fr)\me^{im(\phi-\theta)} \chi(\theta) \psi_0(r,\theta) .
\end{align}
The integral over $\phi$ gives $\delta_{nm}$, then we have
\begin{align}
E&=\frac12\int\md\theta' \int\md\theta \int r\md r \chi^*(\theta') \psi_0^\dag(r,\theta')
\sum_n J_n^2(k_Fr)\me^{in(\theta'-\theta)}
\left[-i\partial_\theta \w_0(\theta) \psi_0(r,\theta) \chi(\theta)\right] \nn\\
&\qquad\qquad\qquad\qquad +\left[i\partial_{\theta'} \w_0(\theta') \chi^*(\theta') \psi_0^\dag(r,\theta') \right] \sum_n J_n^2(k_Fr)\me^{in(\theta'-\theta)}
\chi(\theta) \psi_0(r,\theta) .
\end{align}
Note $\psi_0(r,\theta)$ is a slowly varying function of $r$. The integration at small $r$ is suppressed by the Jacobian $r$. On the other hand, the Bessel function is not square-integrable. These considerations suggest the main contribution to the integration is from $k_F r\gg 1$ (and $r<\xi$), where we can safely approximate $J_n(k_Fr)$ by
\begin{align}
J_n^2(k_Fr)\sim \frac{2}{\pi k_Fr}\cos^2\left(k_Fr-\frac{n\pi}{2}-\frac{\pi}{4}\right)\sim \frac{1}{\pi k_Fr} ,
\end{align}
which is $n$-independent since the cosine square is a fast oscillation term and thus replaced by $1/2$ under the $r$-integral. Then the summation over $n$ gives $\delta(\theta-\theta')$ such that the integral over $\theta'$ can be completed, giving
\begin{align}
E&=\frac{1}{2\pi k_F}\int\md\theta \int \md r \chi^*(\theta) \psi_0^\dag(r,\theta)
\left[-i\partial_\theta \w_0(\theta) \psi_0(r,\theta) \chi(\theta)\right] \nn\\
&\qquad\qquad\qquad\qquad +\left[i\partial_{\theta} \w_0(\theta) \chi^*(\theta) \psi_0^\dag(r,\theta) \right]
\chi(\theta) \psi_0(r,\theta) .
\end{align}
The partial derivative $\partial_\theta$ can act on $\w_0(\theta)$, $\psi_0(r,\theta)$, or $\chi(\theta)$. But only the last one gives a real contribution to $E$ and survives, leading to
\begin{align}
E&=\frac{1}{2\pi k_F}\int\md\theta  \left[\int \md r  \psi_0^\dag(r,\theta)  \psi_0(r,\theta) \right]
\left\{\chi^*(\theta) \w_0(\theta) [-i\partial_\theta \chi(\theta)] + [i\partial_\theta\chi^*(\theta)]\w_0(\theta)\chi(\theta) \right\} .
\end{align}
The integral over $r$ in the square bracket is $1/2$ due to the normalization condition $$\int_{-\infty}^{\infty}\md s \psi_0^\dag(s,\theta) \psi_0(s,\theta)=1,$$
and we get
\begin{align}
E=\frac{1}{4\pi k_F} \int\md\theta \chi^*(\theta) \w_0(\theta) [-i\partial_\theta  \chi(\theta)] + [i\partial_\theta\chi^*(\theta)]\w_0(\theta)\chi(\theta).
\end{align}
By variation of $\chi^*(\theta)$ under the normalization condition $\int\md\theta|\chi(\theta)|^2=1$, we obtain the effective Schr\"odinger equation
\begin{align}
\frac12\{-i\partial_\theta,\w_0(\theta)\}\chi_n(\theta) = E_n \chi_n(\theta).
\end{align}
where $\{\cdot,\cdot\}$ is the anticommutation. This is Eq.~\ref{eq:Heff} in the main text.

\section{Anisotropic Fermi surface}\label{sec:aniso}
Next, we generalize the variational theory to the case of anisotropic Fermi surface. Now $\theta$ is not a good coordinate to label the Fermi momentum $\0k_F$. Instead, we choose the curve length $\ell$ along the Fermi surface as the generalized coordinate, and all $v_F(\ell)$, $k_F(\ell)$, $\theta(\ell)$ are periodic functions of $\ell$. For simplicity, in the following, we will not write their $\theta$-dependence explicitly where applicable.
Another difference from the above isotropic Fermi surface case is the momentum integral on the Fermi surface should be replaced by the integral over $\ell$,
\begin{align}
\iint\md k_x\md k_y\delta(\varepsilon_k) \cdots = \int \md \ell \md q \delta(v_Fq) \cdots = \int \frac{\md\ell}{v_F} \cdots .
\end{align}
Therefore, our ansatz for $\Psi(r,\phi)$ becomes
\begin{align}
\Psi(r,\phi)=\oint\md\ell \chi(\ell)\me^{ik_Fr\cos(\phi-\theta)}\psi_0(r,\theta),
\end{align}
where $v_F(\ell)$ is absorbed in $\chi(\ell)$. This is Eq.~\ref{eq:ansatz2} in the main text.

\begin{figure}[h]
\centering
\includegraphics[width=0.2\linewidth]{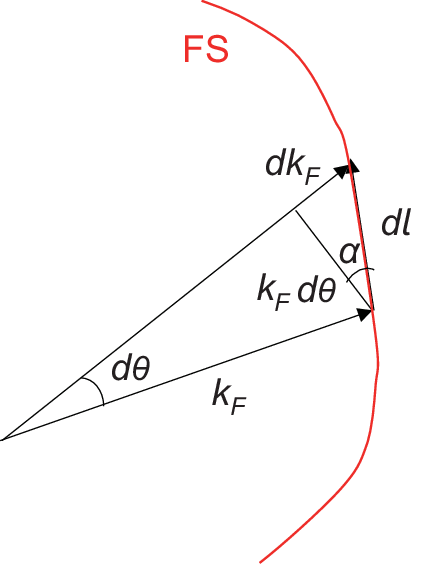}
\caption{ \label{fig:FS-scheme}
Scheme of a segment of the anisotropic Fermi surface contour.
}
\end{figure}

In the following, we use the same strategy as the isotropic case. The main difference is when acting on $\me^{ik_F(\ell)r\cos[\phi-\theta(\ell)]}$, $k_Fb$ should now be replaced by $-ik_F\frac{\md}{\md\ell}$ which can be checked as follows:
\begin{align}
k_Fb\me^{ik_Fr\cos(\phi-\theta)}&=k_Fr\sin(\phi-\theta_v) \me^{ik_Fr\cos(\phi-\theta)} \qquad\text{($\theta_v$ is the angle of $\0v_F$)}\nn\\
&=k_Fr\sin(\phi-\theta+\alpha) \me^{ik_Fr\cos(\phi-\theta)} \qquad(\alpha=\theta-\theta_v) \nn\\
&=k_Fr[\sin(\phi-\theta)\cos\alpha + \cos(\phi-\theta)\sin\alpha] \me^{ik_Fr\cos(\phi-\theta)} \nn\\
&=-i\left(\cos\alpha \frac{\partial}{\partial \theta} + k_F\sin\alpha \frac{\partial}{\partial k_F}\right) \me^{ik_Fr\cos(\phi-\theta)} \nn\\
&=-ik_F\frac{\md}{\md\ell}\me^{ik_F(\ell)r\cos[\phi-\theta(\ell)]},
\end{align}
where we have used the relations $\cos\alpha=k_F\md\theta/\md\ell$ and $\sin\alpha=\md k_F/\md \ell$, which can be seen in Fig.\ref{fig:FS-scheme}. The substitution of $b\to-i\frac{\md}{\md\ell}$ implies that now \textbf{\textit{$\bm{b}$ and $\bm{\ell}$ are a pair of conjugate variables}}. After the substitution, we have
\begin{align}
E&=\frac12\int\md\ell' \int\md\ell \int \md^2\0r \chi^*(\ell') \psi_0^\dag(r,\theta') \me^{-ik_F'r\cos(\phi-\theta')}  \psi_0(r,\theta)\chi(\ell) \w_0(\theta) k_F\left[i\partial_\ell \me^{ik_Fr\cos(\phi-\theta)} \right] \nn\\
&\qquad\qquad\qquad\qquad +\left[-i\partial_{\ell'}  \me^{ik_F'r\cos(\phi-\theta')} \right] k_F' \w_0(\theta') \chi^*(\ell') \psi_0^\dag(r,\theta')\me^{ik_Fr\cos(\phi-\theta)} \psi_0(r,\theta) \chi(\ell) \nn\\
&=\frac12\int\md\ell' \int\md\ell \int \md^2\0r \chi^*(\ell') \psi_0^\dag(r,\theta') \me^{-ik_F'r\cos(\phi-\theta')} \me^{ik_Fr\cos(\phi-\theta)} \left[-i\partial_\ell k_F \w_0(\theta) \psi_0(r,\theta) \chi(\ell)\right] \nn\\
&\qquad\qquad\qquad\qquad +\left[i\partial_{\ell'} k_F' \w_0(\theta') \chi^*(\ell') \psi_0^\dag(r,\theta') \right] \me^{-ik_F'r\cos(\phi-\theta')} \me^{ik_Fr\cos(\phi-\theta)}  \chi(\ell) \psi_0(r,\theta) ,
\end{align}
where $v_F'$, $k_F'$, $\theta'$ are functions of $\ell'$. We have replaced $\frac{\md}{\md\ell}$ by $\partial_\ell$ for short.
Next, we expand $\me^{ik_Fr\cos(\phi-\theta)}$ according to Eq.~\ref{eq:bessel-expansion}, after the integration over $\phi$, we obtain
 \begin{align}
E&=\frac12\int\md\ell' \int\md\ell \int r\md r {\chi^*(\ell')} \psi_0^\dag(r,\theta')
\sum_n J_n(k_F'r)J_n(k_Fr)\me^{in(\theta'-\theta)}
\left[-i\partial_\ell k_F \w_0(\theta) \psi_0(r,\theta) {\chi(\ell)}\right] \nn\\
&\qquad\qquad\qquad\qquad +\left[i\partial_{\ell'} k_F' \w_0(\theta') {\chi^*(\ell')} \psi_0^\dag(r,\theta') \right] \sum_n J_n(k_F'r) J_n(k_Fr)\me^{in(\theta'-\theta)}
{\chi(\ell)} \psi_0(r,\theta) .
\end{align}
Since $\psi_0$ is a slowly varying function of $r$, we use the asymptotic behavior of $J_n$ to write
\begin{align}
J_n(k_F'r)J_n(k_Fr)&\sim \sqrt{\frac{2}{\pi k_F'r}} \sqrt{\frac{2}{\pi k_Fr}} \cos \left(k_F'r-\frac{n\pi}{2}-\frac{\pi}{4}\right) \cos \left(k_Fr-\frac{n\pi}{2}-\frac{\pi}{4}\right) \nn\\
&\sim \frac{1}{\pi k_Fr } \delta_{k_Fk_F'} ,
\end{align}
where the fast oscillation behavior under the $r$-integral gives $\delta_{k_Fk_F'}$, which is a dimensionless Kronecker delta function since the following $n$-summation imposes $\theta(\ell)= \theta(\ell')$ so that there are only a finite number of $k_F(\ell)$ sharing the same $\theta$. Then the summation over $n$ and integral over $\ell'$ yields
\begin{align}
E&=\frac{1}{2\pi }\int \frac{\md\ell}{k_F(\md\theta/\md\ell)} \int \md r {\chi^*(\ell)} \psi_0^\dag(r,\theta)
\left[-i\partial_\ell k_F \w_0(\theta) \psi_0(r,\theta) {\chi(\ell)}\right] \nn\\
&\qquad\qquad\qquad\qquad\qquad +\left[i\partial_{\ell} k_F \w_0(\theta) {\chi^*(\ell)} \psi_0^\dag(r,\theta) \right]
{\chi(\ell)} \psi_0(r,\theta) .
\end{align}
Similarly to the isotropic case in the previous section, the partial derivative $\partial_\ell$ only needs to act on $\chi(\ell)$, and using the relation $\cos\alpha=k_F\md\theta/\md\ell$, we have
\begin{align}
E&=\frac{1}{2\pi}\int\frac{\md\ell}{\cos\alpha} \underbrace{\left[\int \md r  \psi_0^\dag(r,\theta) \psi_0(r,\theta) \right]}_{1/2}
\left\{ {\chi^*(\ell)} k_F\w_0(\theta) \left[-i\partial_\ell {\chi(\ell)}\right] +
\left[i\partial_\ell{\chi^*(\ell)}\right] k_F\w_0(\theta) {\chi(\ell)}
\right\} \nn\\
&=\frac{1}{4\pi} \int\frac{\md\ell}{\cos\alpha} \left\{ {\chi^*(\ell)} k_F\w_0(\theta) \left[-i\partial_\ell {\chi(\ell)} \right] +
\left[i\partial_\ell{\chi^*(\ell)}\right] k_F\w_0(\theta) {\chi(\ell)} \right\} .
\end{align}
By variation with respect to $\chi^*(\ell)$ under the normalization condition (Eq.~[14] in the main text)
\begin{align}
\oint\frac{\md\ell}{\cos\alpha}\chi^*(\ell)\chi(\ell)=1,
\end{align}
we finally obtain
\begin{align}
\frac{1}{2}\left\{ -i\frac{\md}{\md\ell}, k_F(\ell)\w_0(\ell) \right\}{\chi}_n(\ell)=E_n{\chi}_n(\ell) .
\end{align}
which is Eq.~\ref{eq:Heff2} in the main text.

\end{widetext}


\end{document}